\documentclass[12pt,superscriptaddress,aps,pra,onecolumn]{revtex4-1}
\usepackage{dcolumn}
\usepackage{amsmath, amssymb,graphicx,comment,bm,color}
\usepackage{multirow}
\usepackage{booktabs}

\usepackage{multirow}
\usepackage{comment}
\usepackage{xcolor}

\begin{document}
\title{Dipole dynamics in the point vortex model}
\author{Karl Lydon}
\email{lydonk@aston.ac.uk}
\affiliation{Department of Mathematics, College of Engineering and Physical Sciences, Aston University, Aston Triangle, Birmingham, B4 7ET, UK}

\author{Sergey V. Nazarenko} 
\email{sergey.nazarenko@inphyni.cnrs.fr}
\affiliation{Institut de Physique de Nice, Universit\'e C\^ote d'Azur CNRS - UMR 7010, Parc Valrose, 06108 Nice, France}

\author{Jason Laurie}
\email{j.laurie@aston.ac.uk}
\affiliation{Department of Mathematics, College of Engineering and Physical Sciences, Aston University, Aston Triangle, Birmingham, B4 7ET, UK}

\date{\today}

\begin{abstract}
\noindent At the very heart of turbulent fluid flows are many interacting vortices that produce a chaotic and seemingly unpredictable velocity field. Gaining new insight into the complex motion of vortices and how they can lead to topological changes of flows is of fundamental importance in our strive to understand turbulence. Our aim is form an understanding of vortex interactions by investigating the dynamics of point vortex dipoles interacting with a hierarchy of vortex structures using the idealized point vortex model. Motivated by its close analogy to the dynamics of quantum vortices in Bose-Einstein condensates, we present new results on dipole size evolution, stability properties of vortex clusters, and the role of dipole-cluster interactions in turbulent mixing in 2D quantum turbulence. In particular, we discover a mechanism of rapid cluster disintegration analogous to a time-reversed self-similar vortex collapse solution.
\end{abstract}


%
%
%
%
%

\maketitle 
\section{Introduction}\label{sec:intro}

The point vortex model is an Hamiltonian system arising from the consideration of infinitesimally small patches of constant vorticity known as point vortices, that evolve in a self-prescribed irrotational flow by inverting the curl operator using the Biot-Savart law~\cite{kirchhoff_vorlesungen_1876}. A system of point vortices constitutes a weak solution of the 2D Euler equations~\cite{saffman_difficulties_1986}, underlining the importance of point vortices to the study of classical 2D turbulence~\cite{boffetta_two-dimensional_2003}. The model's usefulness to describe ideal 2D flows arises from the fact that vortices in 2D behave like point-like objects advected by the resulting velocity field. Fundamentally, the vortex gas dynamics of the point vortex system can be characterized by a series of scattering and collisions between vortex structures. Subsequently, the point vortex model has been used extensively to model vortex dynamics of 2D flows~\cite{novikov_dynamics_1975,aref_motion_1979,aref_point_1988,aref_integrable_1980}, used in kinetic theories for equilibrium 2D turbulence to understand long time equilibrium states~\cite{nazarenko_kinetic_1992,chavanis_kinetic_2001,chavanis_kinetic_2007}, or been the focus of studies around the possible creation of an Onsager condensate~\cite{onsager_statistical_1949,kiessling_onsagers_2012, salman_long-range_2016}, where same-signed point vortices cluster leading to negative temperature states, in analogy to spectral condensation in classical 2D turbulence~\cite{sommeria_experimental_1986,shats_spectral_2005}.  

The most natural system to compare with the point-vortex model is that of quantum turbulence arising in Bose-Einstein condensates. Here quantum mechanical effects lead to turbulence being a tangle of quasi-1D quantum vortices each identical in structure, with fixed circulation in units of the quantum circulation $\kappa$ and extremely thin vortex cores of the order of $\sim 1\mu m$. In 2D and quasi-2D realizations of quantum turbulence one observes point vortex-like dynamics in a uniform condensate with sound~\cite{nazarenko_freely_2007}. This system has been routinely~\cite{galantucci_turbulent_2011,billam_onsager-kraichnan_2014,reeves_inverse_2013,reeves_enstrophy_2017} modeled using an Euler-like description provided by the point vortex model with the added rule of vortex annihilation to mimic the role of the so-called quantum pressure. It is thought that the inclusion of the vortex annihilation process is important~\cite{salman_long-range_2016} in leading to the long-term formation of an Onsager condensate. This is still an immensely important and open problem within the fluid dynamics and statistical mechanic communities~\cite{montgomery_statistical_1974,eyink_onsager_2006}. A natural question is what are the fundamental processes that lead to mutual approaches of oppositely signed vortices resulting in vortex annihilation in these systems?

In this article, we argue that vortex dipoles are fundamental to this picture. In high temperature states it is conjectured that the system is composed of a sea of tightly formed dipoles that are propagating quickly through the system, while at low temperatures, the system tends to orientate into large-scale clusters of same-signed vortices, or an Onsager condensate. At temperatures in between, we expect to have a mixed state, composed of both vortex clusters and dipoles, with dipoles acting as the high temperature component or noise in the system. As outlined by Salman and Maestrini~\cite{salman_long-range_2016}, the process of vortex annihilation between vortex and anti-vortex pairs is a key mechanism for pushing the system towards lower temperatures. Such a vortex annihilation only occurs if the vortices in a vortex/anti-vortex pair (or dipole) are sufficiently close together, of the order of the vortex core radius, meaning that they are either already tightly grouped or have been dynamically pushed together. 

Dipoles appear amongst all temperature states of larger vortex systems and are the structures that perform the majority of the turbulent mixing due to their fast propagation across the system scattering and colliding with other vortex structures, this means that dipoles are statistically more likely to be involved in collisions and to cause topological changes of the vortex state. 

The structure of this article is as follows: in section~\ref{subsec:point vortex-model} we introduce the 2D point vortex model in the Hamiltonian formulation first described by Kirchhoff~\cite{kirchhoff_vorlesungen_1876} and set out the core mathematical formalism used in our study.  In section~\ref{sec:three_vortex} we briefly review the seminal work of Aref~\cite{aref_motion_1979} on three vortex interactions involving the scattering of a dipole via a third isolated vortex before extending his study to consider the dipole periapsis during the scattering process. In section~\ref{sec:four_vortex} we broaden our study to the analysis of the integrable and chaotic four vortex interaction in dipole-dipole collisions  first studied by Aref and collaborators~\cite{aref_integrable_1980,eckhardt_irregular_1988,eckhardt_integrable_1988} to examine the non-trivial properties of the dipole dynamics during evolution, before finally in section~\ref{sec:dipole-cluster} we investigate the interaction of a dipole with a $m$-vortex cluster and compare the dynamics to those of a general three vortex interaction of a dipole and an isolated vortex of circulation $m\kappa$ with $m$ an integer such that $m\geq 2$ and $\kappa$ the base circulation used in the point vortex model. Section~\ref{sec:conclusion} summarizes our results and give some exciting perspectives on the future direction of this work.

\subsection{The Point Vortex Model}\label{subsec:point vortex-model}

\noindent The point vortex model is a system of $N$ point particles of constant circulation that each generate a corresponding velocity field computed by inverting the curl operator in the vorticity-velocity relation using the Biot-Savart law. The system evolves according to the collective velocity field generated by all $N$ point vortices. This system can be expressed in Hamiltonian form with the position ${\bf x}_i=( x_i,y_i )$ of each point vortex, labeled by the index $i$, evolved through the point vortex evolution equations $\kappa_{i}\dot{x}_{i}=  \partial H/\partial y_{i}, \kappa_{i}\dot{y}_{i}=-\partial H/\partial x_{i}$, with Hamiltonian $H$ given by

\begin{equation}\label{eq:ham}
	H=  -\frac{1}{2\pi}\sum_{i=1}^{N}\sum_{j=1, j<i}^{N}\kappa_{i}\kappa_{j}\ln\left(l_{ij}\right),
\end{equation}

\noindent where $\kappa_{i}$ is the circulation of point vortex $i$, $x_{ij}=x_{i}-x_{j}$ and $y_{ij}=y_{i}-y_{j}$ are the $x$ and $y$ distances between two point vortices labeled as $i$ and $j$, with the square of their distance defined by $l_{ij}^{2}=x_{ij}^{2}+y_{ij}^{2}$. Here we denote the temporal derivative of a function $f$ by $\dot{f}={\rm d}f/{\rm d}t$. Consequently, the corresponding equations of motion of an individual vortex position can be expressed as

\begin{equation}\label{eq:pv}
	\dot{x}_{i}=-\frac{1}{2\pi}\sum_{j=1, j\neq i}^{N}\frac{\kappa_{j}y_{ij}}{l_{ij}^{2}},\qquad  \dot{y}_{i}=\frac{1}{2\pi}\sum_{j=1, j\neq i}^{N}\frac{\kappa_{j}x_{ij}}{l_{ij}^{2}}.
\end{equation}

  In the infinite domain, the point vortex system possesses a set of symmetries; namely, spatial translations in $x$ and $y$, and any arbitrary rotation around a fixed point, that will not lead to any change in the overall dynamics of the system. These inherent symmetries can be mapped onto conservation laws through Noether's theorem~\cite{goldstein_classical_1980} which relates the respective symmetries to the conservation of the linear momentum ${\bf P}$ and angular momentum $M$ respectively:
\begin{equation}\label{eq:momentum}
{\bf P}=( P_x,P_y )=  \left( \sum_{i=1}^{N}\kappa_{i}x_{i},\sum_{i=1}^{N}\kappa_{i}y_{i}\right ), \qquad M=  \sum_{i=1}^{N}\kappa_{i}\left(x_{i}^{2}+y_{i}^{2}\right).
\end{equation}

\noindent The conservation of the linear momentum imply that the central point of circulation $\mathbf{x}_{\Gamma}=( x_{\Gamma},y_{\Gamma} )= (1/\Gamma_{\rm tot})\left( P_x,P_y\right )$, becomes a fixed point of the vortex dynamics with the total circulation $\Gamma_{\rm tot}$ of the point vortex system defined by $\Gamma_{\rm tot}= \sum_{i=1}^{N}\kappa_{i}$.

The equations of motion~(\ref{eq:pv}) constitute a $2N$ dynamical system supported by four conservation laws: the Hamiltonian~(\ref{eq:ham}), the two linear momentum coordinates, and the angular momentum~(\ref{eq:momentum}), meaning that the point vortex system in the infinite domain then has a total of $2N-4$ degrees of freedom.

For small $N$, it is often convenient to study the point vortex model in a description where the dynamics of the system is specified by the vortex separation distances $l_{ij}$ between pairs of point vortices and not by the specific frame coordinates. Re-expressing~(\ref{eq:pv}) using the conservation laws leads to an evolution equation for the square of the separation distances $l_{ij}$~\cite{aref_motion_1979}:

\begin{equation}\label{eq:l-dyn}
	\frac{dl_{ij}^{2}}{dt}=  \frac{2}{\pi}\sum_{k=1 , k\neq i ,  k \neq j}^{N}\kappa_{k}\epsilon_{ijk}A_{ijk}\left(\frac{1}{l_{jk}^{2}}-\frac{1}{l_{ki}^{2}}\right),
\end{equation}

\noindent where $\epsilon_{ijk}$ is the Levi-Civita symbol that indicates the orientation of the triangle spanned by the three vortices $i,j,k$, taking values $-1$ if $i,j,k$ are orientated in a clockwise fashion and $+1$ if anti-clockwise. $A_{ijk}$ represents the area of the triangle created by the three vortices $i, j, k$ which can be computed by Heron's formula using the known lengths of the triangle sides $A_{ijk} =\sqrt{p(p-l_{ij})(p-l_{ik})(p-l_{jk})}$ with $p=(l_{ij}+l_{ik}+l_{jk})/2$ representing the semi-perimeter of the triangle spanned by the vortices. We can construct a coordinate-free conservation law from the original linear and angular momenta, which we denote as $R$ 
\begin{equation}\label{eq:R}
	R= \Gamma_{\rm tot} M-|{\bf P}|^{2} = \sum_{i=1}^{N}\sum_{j=1, j<i}^{N}\kappa_{i}\kappa_{j}l_{ij}^{2}.
\end{equation}

\noindent The separation length description yields a dynamical system of $N(N-1)/2$ degrees of freedom with two conservation laws:~(\ref{eq:ham}) and~(\ref{eq:R}), leading to $N(N-1)/2 -2$ overall degrees of freedom. 

\subsection{Numerical Simulations of the Point Vortex Model}

\noindent In all of the following work, we perform numerical simulation of the point vortex model~(\ref{eq:pv}) in a infinite 2D domain using an adaptive fourth-order Dormand-Prince Runge–Kutta method, details of which can be found in Ref.~\cite{press_numerical_2009}. All simulations are performed using an adaptive time-stepping scheme because we found that it was essential for the conservation of the Hamiltonian~(\ref{eq:ham}) and momentum~(\ref{eq:momentum}). In all our simulations, we set the point vortex circulations $\kappa_i=\pm 1$, and a maximum time-step of $\Delta t \leq 10^{-3}$. In all simulations we compute the relative error of conserved quantities to ensure an acceptable level of accuracy is maintained, we define the relative error for the Hamiltonian at time $t$ as $|H(t)-H(0)|/H(0)$ where $H(t)$ is the Hamiltonian at time $t$ and $H(0)$ is the computed Hamiltonian from the initial condition.

\section{Three Vortex System: Dipole-Vortex Collisions}\label{sec:three_vortex}

\noindent The three vortex system has been shown to be integrable, with the phase space dynamics characterized by Novikov~\cite{novikov_dynamics_1975} and Aref~\cite{aref_motion_1979}. Our interest is in the examination of the special case in which a dipole interacts with a third isolated point vortex. In terms of turbulent flows composed of a dilute system of point vortices, we expect the dipole-vortex collision to be the most common interaction and hence the most important type of interaction in terms of vortex mixing. The dipole-vortex scattering problem was originally studied by Aref in his seminal work~\cite{aref_motion_1979}, where Aref applied the phase space formalism of Novikov~\cite{novikov_dynamics_1975} to first characterize regions of direct and exchange scattering, and then determine analytically the dipole scattering angle. Apart from a brief review of the mathematical formalism used by Novikov and Aref in the present section, we leave the precise details of calculations of the dipole scattering angles (with an unseen mistake corrected) to the appendix. 

\begin{figure}[htp!]
\begin{center}
\includegraphics[width=0.7\columnwidth]{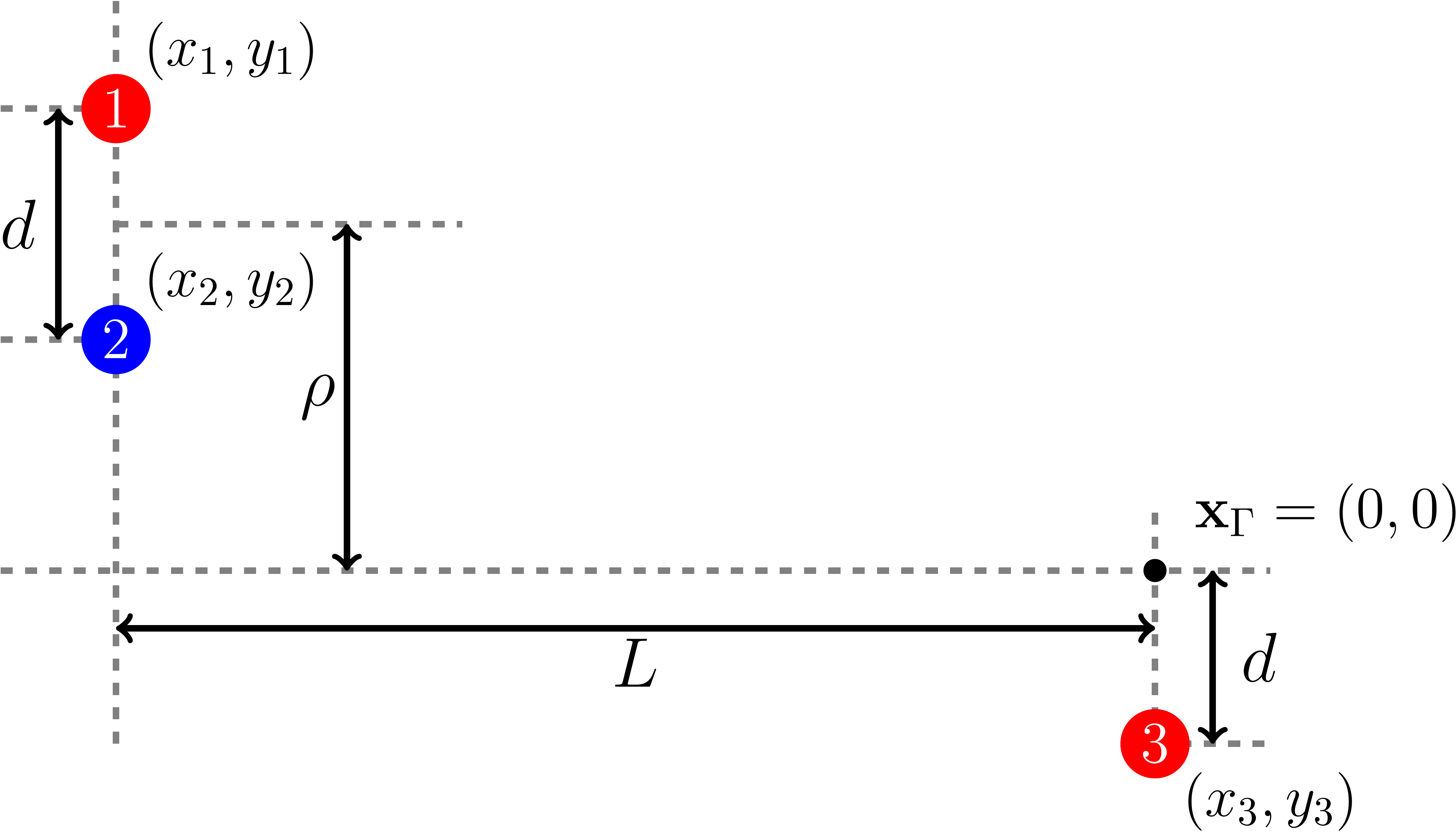}
\end{center}
\caption{The initial setup of the dipole-vortex interaction. The red circles indicate positions of the positive circulation $\kappa$ point vortices, while the blue circle indicates the negative circulation $-\kappa$ point vortex. 
\label{fig:1_three_vortex_setup}}
\end{figure}

The dipole-vortex scattering process can be fully characterized by the setup depicted in figure~\ref{fig:1_three_vortex_setup}; we introduce a vertically orientated vortex dipole composed of oppositely signed vortices labeled $1$ and $2$ with circulations $\kappa_1=\kappa>0$ and $\kappa_2=-\kappa$ respectively that are situated a horizontal distance $L$ from a third isolated point vortex $3$ of positive circulation $\kappa_3=\kappa$. We limit ourselves to point vortices of identical circulation magnitude to keep our analogy with quantum vortices in Bose-Einstein condensates. Vortices $1$ and $2$ of the vortex dipole are a vertical distance $d$ from each other. We define an impact parameter $\rho$ that quantifies the vertical distance of the midpoint of the vortex dipole to the center of circulation ${\bf x}_{\Gamma}=(0,0)$ which we have set as the origin of our coordinate frame. This results in the following values of the conserved quantities: Hamiltonian $ H=  -(\kappa^2/2\pi)\ln\left[ (l_{13})/(l_{12}l_{23})\right ] \stackrel[L\to\infty]{}{\rightarrow} (\kappa^2/4\pi)\ln\left( d^2\right)$ and the linear and angular momenta result in
$ {\bf P} = \left( 0,0 \right ), M= \kappa(d^2+2\rho d ) $ leading to the conversed value of $R = \kappa^2(d^{2}+2\rho d)$. Also as the three lengths between any two vortices form a triangle, we must also have the following geometric constraints $l_{12} \leq l_{13}+l_{23}$, $l_{13} \leq l_{12}+l_{23}$, and $l_{23} \leq l_{12}+l_{13}$. We consider the dipole-vortex interaction process in the limits of $d,\rho\ll L$ to ensure that the initial propagation of the dipole is unaffected by the presence of the third isolated vortex. The phase space analysis of Aref shows there are two types of interaction possible; {\it direct scattering} for large $|\rho|$ where the dipole propagates past vortex $2$ with only a deflection in propagation angle occurring, and {\it exchange scattering} for small  $|\rho|$ where the dipole positive vortex and isolated vortex exchange, disturbing the initial dipole structure.

We denote the exact moment of any exchange scattering as \textit{the exchange point} in this case, the exchange point is defined when the two distances $l_{12}=l_{23}$. Furthermore, we can also define the critical point of the interaction as the moment during the interaction where the dipole size is at a local minimum or maximum, of particular interest are the inter-vortex lengths at this point, which we represent as $l_{ij}^*$. Note that because of the simple nature of the three vortex collision, the exchange point and the critical point occur simultaneously, but in more complicated collisions involving more than three point vortices this may not necessarily be the case. Similarly for direct scattering, the critical point of~(\ref{eq:l-dyn})  arises when all three point vortices become collinear with the value of the area of the triangle spanned by the three vortices $A_{123}$ vanishing.  At this moment, one either has $l_{13} = l_{12} + l_{23}$  (for $\rho >0$) or $l_{23} = l_{12}+l_{13}$ (for $\rho < 0$). At the interface between the direct and exchange scattering regions we expect that the vortices will be trapped in a bounded state of constant rotation as was shown previously by Aref~\cite{aref_motion_1979}.

Following the works of~\cite{novikov_dynamics_1975, aref_motion_1979}, we introduce dimensionless variables $b_i$ for $i=1,2,3$ defined in terms of the square of the inter-vortex lengths $ b_1=l^2_{23}/(\kappa C), b_2=-l^2_{13}/(\kappa C), b_3=l^2_{12}/( \kappa C )$, where $C$ is a constant of motion derived from the previous conserved quantity $R$, similarly we attain the conserved quantity $\theta$ from  the Hamiltonian $H$: $ \theta = |b_2|/(b_1b_3) =  \kappa|C|\exp\left(-4\pi H/\kappa^2\right)$. From this and the vortex separation equation of motion~(\ref{eq:l-dyn}), the equation of motion for $b_2$ is found in order to encapsulate the interaction as a whole

\begin{equation}\label{eq:b2}
	\dot{b}_2 = \begin{cases}
	\pm\frac{\sqrt{\theta}}{C\pi b_2}\sqrt{\left(\alpha-b_2\right)\left(\beta-b_2\right)\left(\gamma-b_2\right)} & \text{for $C > 0$}, \\ \pm\frac{\sqrt{\theta}}{C\pi b_2}\sqrt{\left(b_2-\bar{\alpha}\right)\left(b_2-\bar{\beta}\right)\left(b_2-\bar{\gamma}\right)}
 & \text{for $C < 0$},
 \end{cases}
\end{equation}

\noindent with $\alpha, \beta, \gamma$ and $\bar{\alpha}, \bar{\beta}, \bar{\gamma}$ all functions of $\theta$, giving the roots of the equation of motion $\dot{b}_2$ for the respective values of $C$. We give the full procedure of introducing dimensionless coordinates and their relation to real-space vortex dynamics in the appendix. From consideration of which root is encountered first from the initial state defined at $t\to -\infty$ we can determine the boundary of the scattering regions with respect to parameters $C$ and $\theta$. This is displayed in table~\ref{tab:1_three_vortex_regions}. Note that although the roots $\beta$ and $\bar{\beta}$ also correspond to the condition $l_{12}=l_{23}$, $b_2$ will always reach one of the other two sets of roots first due to our initial vortex setup defined as the dipole situated away at infinity. For the dipole scattering process that we have defined it is only possible to reach these roots in the special cases of $\theta=1/3$ and $\theta=8/3$ when $\beta=\gamma$ and $\bar{\beta}=\bar{\gamma}$ respectively. In these cases the point vortices form a quasi-stable equilateral triangle or collinear structure that exhibits rigid-body rotation about the center of circulation (confirmed by our numerical computations).

\begin{table}[htp!]
\begin{center}
\caption{Parameter ranges of both $\rho, d$ and $\theta$ corresponding to scattering types of the three regions of the three vortex interaction defined in Fig.~\ref{fig:1_three_vortex_setup}. Region II has been split into two sub regions defined by the sign of $C$, although the same scattering type is observed in these regions.}
\setlength{\tabcolsep}{5pt}
\renewcommand{\arraystretch}{1.25}
\begin{tabular}{c|c|c|c|c}
Region & $C$ Range & $\theta$ Range & Impact Parameter Range & Scattering Type\\
\hline
I & $C > 0$ & $1/3 <\theta < \infty $ & $-\infty< \rho/d < -1$ &  Direct\\
IIa & $C > 0$&$0 <\theta < 1/3$ & $ -1 <\rho/d < -1/2$ & Exchange\\
IIb& $C  < 0$& $0 < \theta < 8/3$ & $-1/2 < \rho/d < 7/2$ & Exchange\\
III & $C < 0$&$8/3<\theta < \infty$ & $7/2 < \rho/d < \infty $ &Direct \\ 
\end{tabular}\label{tab:1_three_vortex_regions}
\end{center}
\end{table}

\subsection{Dipole-Vortex Scattering}

\begin{figure}[htp!]
\begin{center}
\includegraphics[width=0.8\textwidth]{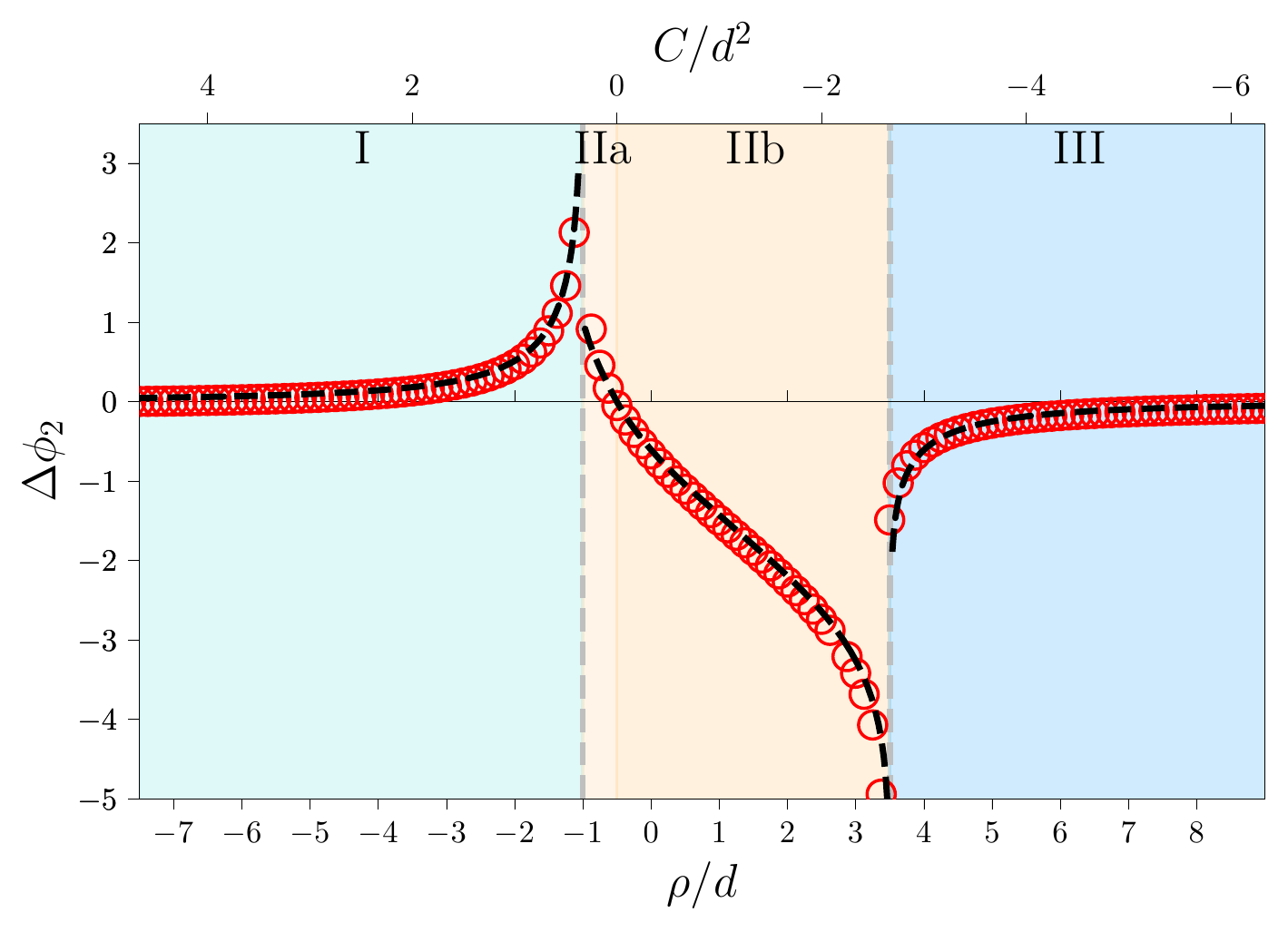}
\caption{Unwinded scattering angle of the negatively signed point vortex in the dipole-vortex interaction with respect to the normalized impact parameter $\rho/d$. Numerical results using the point vortex model~(\ref{eq:pv}) are given by red circles, while the theoretical predictions are plotted as the black dashed curve. Also marked are the regions corresponding to different scattering processes as presented in table~\ref{tab:1_three_vortex_regions}. \label{fig:2_three_vortex_angles}}
\end{center}
\end{figure}

\noindent Aref~\cite{aref_motion_1979} used the formalism presented above to determine analytical results for the dipole scattering angle after interaction with the third isolated point vortex. The analytical results of Aref are re-derived in the appendix with a small mistake corrected, and are plotted in figure~\ref{fig:2_three_vortex_angles} in comparison with numerical results. All numerical simulations in the dipole-vortex case are performed such that the relative error of the Hamiltonian is conserved to $10^{-12}$. Our definition of the scattering angle is only defined for modulo $2\pi$, however in figure~\ref{fig:2_three_vortex_angles} we have unraveled the scattering angle to better display the meaning of the asymptotes. Positive values of $\Delta \phi_2$ indicate an anti-clockwise deflection.  We observe excellent agreement between the theoretical results given in the appendix  (black dashed curve) and the numerical data (red circles). For large values of the normalized impact parameter $\rho/d$ we observe minimal deflection of the dipole as expected. As the impact parameter shrinks, corresponding to a more direct, head on, propagation of the dipole towards the isolated vortex we observe two clear asymptotes, one between regions I and IIa  and another between regions IIb and III indicating the boundaries between the direct and exchange scattering regions. The asymptotes correspond to locking of the three vortices into either an equilateral triangle or a quasi-stable collinear structure that undergoes continual rotation - hence the tend towards an infinite scattering angle. Region IIa and IIb correspond to exchange scattering where we observe an almost $-\sinh$ like behavior. We note that figure~\ref{fig:2_three_vortex_angles} slightly differs from what was originally presented in~\cite{aref_motion_1979} (figure 11) due to a sign error made in that paper. These calculations have been corrected and can be found in the appendix.  

\subsection{Dipole Size and the Periapsis} \label{sec:3v-dipole_lengths}

\noindent As well as quantifying the angle of deflection of a dipole with a third vortex, it is also important to examine the dipole size during its propagation. This is an essential piece of dynamical information if one draws a connection of point vortices to quantum vortices in Bose-Einstein condensates because quantum vortices can undergo a process of annihilation  if they proceed to interact within a critical distance to each other. Approximately this occurs at the order of the quantum vortex core radius or healing length $\xi\sim 1\mu m$. Consequently, the process of vortex annihilation leads to the reduction of the number of vortices and the generation of sound in Bose-Einstein condensates~\cite{nazarenko_freely_2007}, which can dramatically change the vortex topology of the turbulence.

Subsequently, in this subsection, we will focus on two main quantities: the periapsis -- a term borrowed from celestial mechanics meaning the distance of closest approach -- and the final dipole size post-interaction. In the case of the dipole-vortex interaction considered here, the final dipole size (post-interaction at $t\to\infty$) will trivially equal $d$ due to energy conservation of the system. The periapsis, can be determined via the roots of~(\ref{eq:b2}), as these correspond to critical points of the differential equation for $\dot{l}^2_{ij}$ for all the inter-vortex distances. For the dipole-vortex interaction the critical points for all the inter-vortex length occur at the same moment $t^*$. The critical points of~(\ref{eq:l-dyn}) will then be local minimum or maximum lengths of the inter-vortex distances. Depending on the interaction region, we can then compute the local minimum or maximum inter-vortex lengths of the formed dipole.

\begin{figure}[htp!]
\begin{center}
\includegraphics[width=0.8\textwidth]{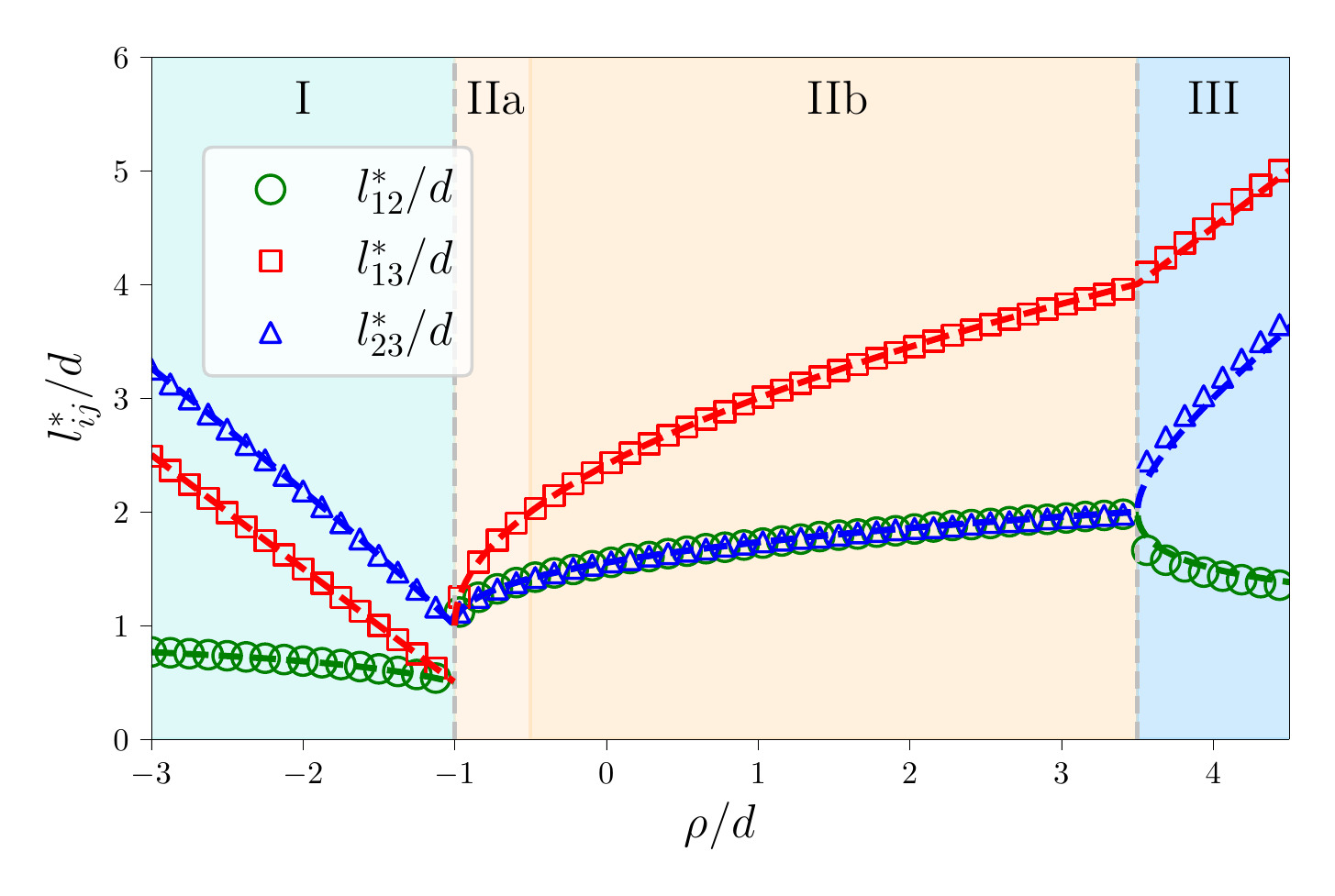}
\caption{Numerical results compared with theoretical predictions of the critical lengths in the dipole-vortex interaction with respect to the normalized impact parameter $\rho/d$. Dashed lines represent the theoretical predictions of each length determined by the critical points of~(\ref{eq:l-dyn}). Markers represent the numerical results for the corresponding lengths as indicated by the legend. Note the asymptotic cases at $\rho/d=-1$ and $\rho/d=7/2$ corresponding to rigid-body motion. \label{fig:3_three_vortex_critical_lengths} }
\end{center}
\end{figure}

We define variables $b_i^*$ representing the critical distance of the $i_{th}$dimensionless variable, simple calculations as found in the appendix can then determine the value of $b_2^*$, the $b$ variable encapsulating the motion, at the periapsis. Once the value of $b^*_2$ is found at the periapsis it is straightforward to determine all the lengths between the vortices using the relationship between $b_1, b_2, b_3$ or equivalently $l_{12}, l_{13}, l_{23}$. Careful consideration of the correct sign in resulting square roots needs to be applied however, but the resulting lengths of the periapsis of the scattering process in the limit of $\rho \ll L$ and $d \ll L$ are given in table~\ref{tab:2_three_vortex_critical_lengths}. We verify the analytical results against those obtained from numerical computations using the point vortex model. Results are displayed in figure~\ref{fig:3_three_vortex_critical_lengths} where excellent agreement is observed partly due to the analytical results given in table~\ref{tab:2_three_vortex_critical_lengths} being $O((\rho/L)^2)$ and $O\left((d/L)^2\right)$ accurate in the limit $\rho, d\ll L$.

\begin{table}[htp!]
\begin{center}
	\caption{Vortex separations at the periapsis of the three vortex scattering setup given in terms of the impact parameter $\rho$ and the initial dipole separation $d$ for each region of interaction. Values are leading order results taken in the limits of initial far dipole separation: $\rho \ll L$ and $d \ll L$.\label{tab:2_three_vortex_critical_lengths}}
\setlength{\tabcolsep}{5pt}
\renewcommand{\arraystretch}{1.5}
\resizebox{\textwidth}{!}{
\begin{tabular}{c|c|c|c}

	Region & $l^*_{12}$ &  $l^*_{13}$ & $l^*_{23}$ \\
\hline
	I & $d\left( \frac{1}{4} + \frac{1}{2}\frac{\rho}{d} - \frac{1}{2}\sqrt{-\frac{7}{4} - 3\frac{\rho}{d} + \frac{\rho^2}{d^2}} \right)$ & $d\left(-\frac{\rho}{d}-\frac{1}{2}\right)$ & $d\left( \frac{1}{4} + \frac{1}{2}\frac{\rho}{d} + \frac{1}{2}\sqrt{-\frac{7}{4} - 3\frac{\rho}{d} + \frac{\rho^2}{d^2}} \right)$ \\
	IIa &	$d\sqrt{1+\sqrt{2\left(1+\frac{\rho}{d}\right)}}$& $d\left(1 + \sqrt{2\left(1+\frac{\rho}{d} \right)} \right)$ & $d\sqrt{1+\sqrt{2\left(1+\frac{\rho}{d}\right)}}$  \\
	IIb & $d\sqrt{1+\sqrt{2\left(1+\frac{\rho}{d}\right)}}$& $d\left(1 + \sqrt{2\left(1+\frac{\rho}{d} \right)} \right)$ & $d\sqrt{1+\sqrt{2\left(1+\frac{\rho}{d}\right)}}$  \\
	III & $d\left( \frac{1}{4} + \frac{1}{2}\frac{\rho}{d} - \frac{1}{2}\sqrt{-\frac{7}{4} - 3\frac{\rho}{d} + \frac{\rho^2}{d^2}}\right)$ & $d\left(\frac{\rho}{d}+\frac{1}{2}\right)$ & $d\left( \frac{1}{4} + \frac{1}{2}\frac{\rho}{d} + \frac{1}{2}\sqrt{-\frac{7}{4} - 3\frac{\rho}{d} + \frac{\rho^2}{d^2}}\right)$ \\ 
\end{tabular}
}
\end{center}
\end{table}

\begin{figure}[htp!]
	\begin{center}
		\includegraphics[width = \textwidth]{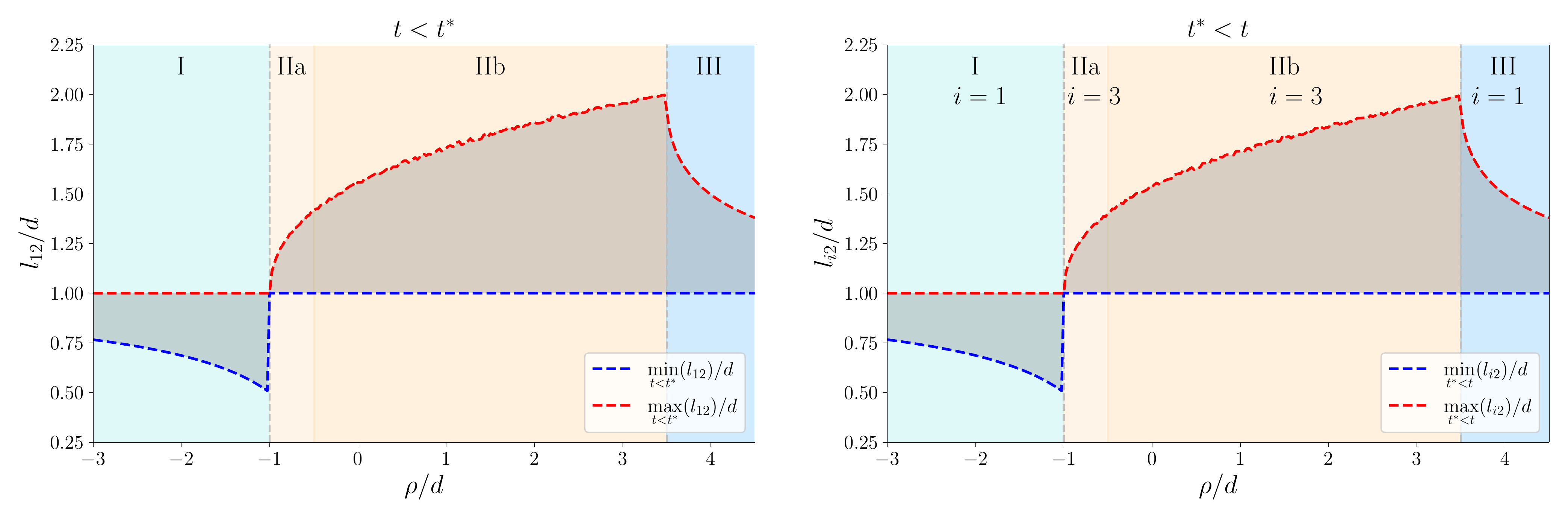}
	\caption{Numerical results of the minimum and maximum size of the propagating vortex dipole pre- and post-interaction with the third isolated vortex. The moment of interaction is defined via the value of $t^*$ that indicates the critical point. In region II, the tracked vortex dipole is that composed of vortex $2$ and $3$. \label{fig:4_three_vortex_dipole_size} }
\end{center}
\end{figure}

In figure~\ref{fig:4_three_vortex_dipole_size} we plot the global minimum and maximum lengths of the vortex dipole pre- and post-interaction defined at the critical point $t^*$. The blue and red dashed lines are numerical measurements, but agree with the theoretical critical inter-vortex distances presented in table~\ref{tab:2_three_vortex_critical_lengths} and verified in figure~\ref{fig:3_three_vortex_critical_lengths}. We observe that the two figures are identical as the critical values of the dipole size are defined either at the initial or end states $=d$ or at the critical point $t^*$. Note that in region II, post-interaction the dipole is composed of vortex $2$ and $3$.

What is noteworthy is that in region I, for negative impact parameters $\rho/d < -1$, the dipole distance shrinks during propagation, with the minimum distance achieved at the periapsis point with size $l_{12}=d/2$ when $\rho = -d$. This implies that the three vortex scattering process can bring dipoles closer together as much as a factor of two (although only temporarily due to energy conservation leading to the post-interaction dipole size returning to $d$). Ultimately, this could be a viable mechanism for vortex annihilation in compressible point vortex like systems such as Bose-Einstein condensates pushing quantum vortex dipoles close enough for vortex annihilation to occur as indicated in~\cite{nazarenko_freely_2007}. In regions IIa and IIb we observe that $l_{12}=l_{23}$ at the critical point confirming the exchange scattering scenario described previously and that the exchange scattering occurs at the periapsis with a widening of the dipole vortex (in this case it corresponds to maximum of the dipole distance during the complete interaction). In region III, a direct scattering interaction for positive impact parameter, the dipole size increases from its initial size $d$ during its interaction before returning back to size $d$. At large impact parameters (both positive and negative) the dipole passes by the third vortex at a significant  distance  meaning any three vortex scattering will be negligible with $l_{12}$ remaining close to the initial value $d$.

\section{Four Vortex System: Dipole-Dipole Collisions}\label{sec:four_vortex}

Here we consider the interaction of two identical dipoles, a specific type of four vortex interaction. The four vortex system is in general a non-integrable system, hence analytical methods can not be applied. However, the dipole-dipole collision has been studied by both Aref and Eckhardt~\cite{eckhardt_integrable_1988-1,eckhardt_integrable_1988,eckhardt_irregular_1988} where special regimes of integrable motion have been identified and theoretical expressions for the dipole scattering angles computed~\cite{eckhardt_integrable_1988-1,eckhardt_integrable_1988}. The integrable cases arise when the linear momentum and the total circulation of the system mutually vanish, corresponding to dipoles forming a parallelogram leading to additional symmetries that enable one to reduce the mathematics of the four vortex system to an integrable three-body problem similar to that discussed in section~\ref{sec:three_vortex}. In the following two subsections we will discuss both the integrable and non-integrable dipole-dipole interactions.

\subsection{Integrable Dipole-Dipole Scattering}
\label{sec:int-4vortex}

\begin{figure}[htp!]
\begin{center}
\includegraphics[width=0.7\textwidth]{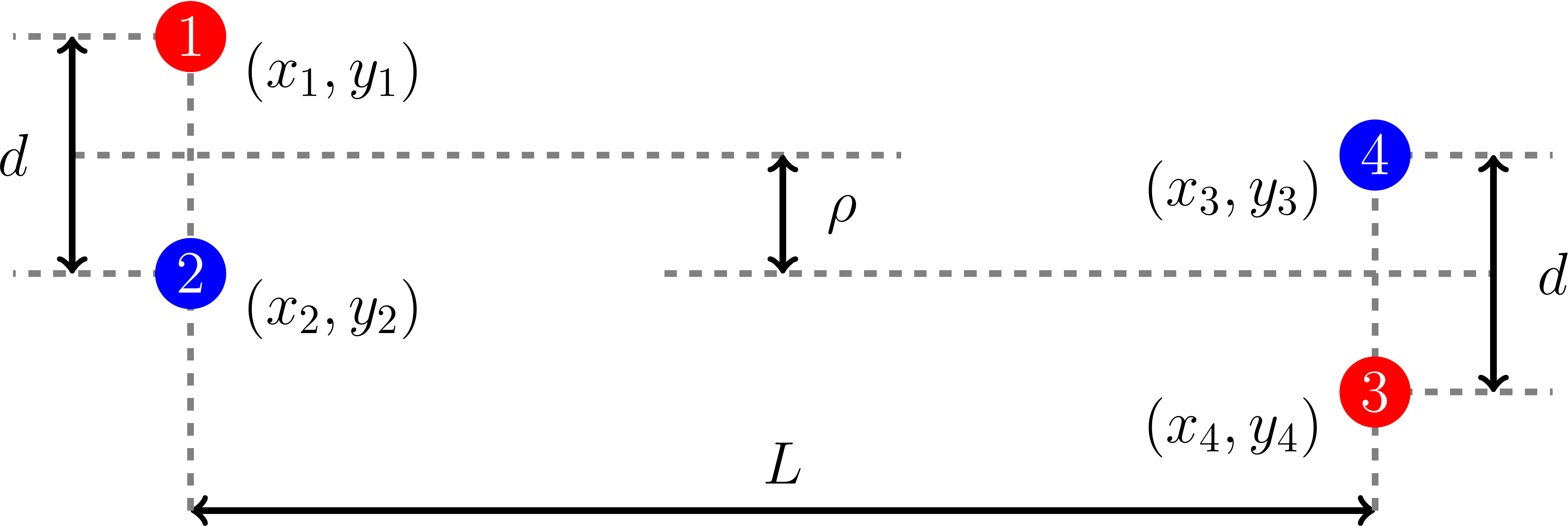}
\caption{Initial setup of the integrable four vortex interaction, with the initial dipole separations defined as $d$, and impact parameter between dipole midpoints represented as $\rho$, and horizontal separation between dipoles $L$. \label{fig:5_four_vortex_integrable_setup}}	
\end{center}
\end{figure}

The integrable dipole-dipole collision corresponds to a configuration where two dipoles propagate along the same geometric axis in the form of a parallelogram. As shown in figure~\ref{fig:5_four_vortex_integrable_setup}. The parallelogram geometry is preserved during evolution leading to an extra geometrical constraint on the motion of the four vortices enabling a mathematical reduction of the system to an effective three body interaction. 

Our initial setup, presented in figure~\ref{fig:5_four_vortex_integrable_setup}, considers two direct facing point vortex dipole each of size $d$ and circulation strengths $\pm \kappa$. The impact parameter $\rho$ characterizes the vertical displacement between the midpoints of the initial dipole positions. Parameter $L$ describes the initial horizontal separation between the two dipoles, and it is in the limit of $\rho, d\ll L$ in which we consider. Indeed, the parallelogram geometry allows us to exclude consideration of the fourth vortex from the system using the relations $l_{34}=l_{12}$, $l_{14}=l_{23}$, and $l_{24}^2=2l_{12}^2+2l_{23}^2-l_{13}^2$ to arrive at a three vortex scattering problem, albeit with a more complicated Hamiltonian taking the form $ H=-(\kappa^2/2\pi)\ln\left[ \left(l_{13}\sqrt{2l_{12}^2+2l_{23}^2-l_{13}^2}\right)/\left(l_{12}^2l_{23}^2\right)\right ] \stackrel[L\to \infty]{}{\rightarrow}(\kappa^2/2\pi)\ln\left(d^2\right)$

In the $L\to\infty$ limit, the Hamiltonian gives twice the value of the Hamiltonian in the three-vortex problem in section~\ref{sec:three_vortex}, as an infinitely isolated vortex contributes no energy to the system. The corresponding momentum for the integrable dipole-dipole interaction vanishes giving ${\bf P}=(0,0)$, $M=0$, and leads to $R=0$ if all four vortices are considered. 

Here, it is appropriate to use an effective momentum $R_{\rm eff}= 2\kappa^2\rho d$ by considering only the contributions to $R$ from the three vortices $1$, $2$, and $3$ as the position of the fourth vortex (say vortex $4$) are slave variables to those of the other three due to the conserved parallelogram symmetry, thus forming the extra conserved quantity $R_{\rm eff}$.   

As in the three vortex case, scattering is encapsulated through the dimensionless variable $b_2$ as these vortices can not form a dipole and so can be used to track the interaction. Hence we introduce dimensionless variables as in the three vortex case, with $C$ defined as in the three vortex case and a slightly different formula for $\theta$ on account of the more complicated Hamiltonian $ \theta = \sqrt{b_2(b_2-6)}/(b_1b_3) = \kappa |C|\exp\left(-2\pi H/\kappa^2\right)$. From consideration of the equation of motion between positive vortices $1$ and $3$ using~(\ref{eq:l-dyn}) with the already discussed geometric constraints and dimensionless $b_i$ variables as in the three vortex case we attain the equation of motion for $b_2$

\begin{equation}\label{eq:4vortex_eq_dimensionless} 
	\dot{b}_2 = \begin{cases}\frac{2\sqrt{\theta}}{C\pi\sqrt{b_2(b_2-6)}}\sqrt{(b_2-\alpha)(b_2-\beta)(b_2-\gamma)} & \text{for $C > 0$}, \\
	 \frac{2\sqrt{\theta}}{C\pi\sqrt{b_2(b_2-6)}}\sqrt{(\bar{\alpha}-b_2)(\bar{\beta}-b_2)(\bar{\gamma}-b_2)} & \text{for $C < 0$.}\\
\end{cases}
\end{equation}

\noindent As in the dipole-vortex interaction we have three sets of roots for $\dot{b}_2$, either $\alpha, \beta, \gamma$ or $\bar{\alpha}, \bar{\beta}, \bar{\gamma}$ depending on what sign the quantity $C$ takes. Note that these are not the same roots as found in the dipole vortex case, one can consult the appendix for full expressions of the roots in this case, as well as the equation of motion derivation and the relation to physical dynamics. From this we give in table~\ref{tab:3_four_vortex_regions} the four regions of interaction and the parameter ranges in which they apply.

\begin{table}[htp!]
\begin{center}
	\caption{Parameter ranges of $C, \theta$, and $\rho/d$ by interaction region and the corresponding scattering types  of the four vortex interaction, similar to \ref{tab:1_three_vortex_regions}. Region II has been split into two sub regions defined by the sign of $C$ with both regions IIa and IIb exhibiting exchange scattering.\label{tab:3_four_vortex_regions}}
\setlength{\tabcolsep}{5pt}
\renewcommand{\arraystretch}{1.25}
\begin{tabular}{c|c|c|c|c}
Region & $C$ Range & $\theta$ Range & Impact Parameter Range & Scattering Type\\
\hline
I & $C > 0$ & $2/3 <\theta $ & $\rho/d < -1$ &  Direct\\
IIa & $C > 0$&$0 <\theta < 2/3$ & $ -1 <\rho/d < 0$ & Exchange\\
IIb& $C < 0$ & $0 < \theta < 2/3$ & $0 < \rho/d < 1$ & Exchange\\
III & $C < 0$&$2/3<\theta $ & $1 < \rho/d $ &Direct\\
\end{tabular}
\end{center}
\end{table}

The boundary case $\theta = 2/3$ corresponds to vortices of the same sign colliding on exactly opposite trajectories, leading to a double Havelock ring configuration~\cite{havelock_stability_1931} whereupon vortices form a rhombus with diagonals of length $d\sqrt{2\sqrt{2}-1}$ and larger diagonal $d\sqrt{2\sqrt{2}+1}$ rotating about the center with constant angular velocity $\dot{\phi}=\pm(\kappa/2\pi d^2)(3+\sqrt{2})$. When $C>0$ at the limit case we have the positive signed vortices colliding along opposite trajectories, it is these vortices that then form the smaller diagonal and we then have anti-clockwise rotation of the rhombus, thus vortices take the negative value of angular velocity. Conversely, in the case that $C<0$ it is the negative vortices that collide on opposite trajectories, which then form the smaller diagonal, and then clockwise rotation is observed. Also, the case when $C=0$ corresponds to two vortex dipoles colliding head on and is analogous to a vortex ring colliding head on with a wall using the technique of images. As is expected in this kind of interaction the exchange scattering leads to new dipoles scattered at right angles to the propagation of the original dipoles.

The scattering angle of a vortex dipole in the integrable dipole-dipole interaction has been found by Eckhardt and Aref~\cite{eckhardt_integrable_1988}, we compare this against numerical simulation data and find total agreement, so the derivation will not be repeated here. As in the dipole-vortex case, all simulations are performed such that the relative error of the Hamiltonian is conserved within $10^{-12}$. 

Now we will examine the evolution of the dipole size in the integrable dipole-dipole collision. By considering the root first reached by $b_2$, we can, by analysis equivalent to that already performed in subsection~\ref{sec:3v-dipole_lengths}, determine the periapsis of the variable $b_2$ across the parameter space. As in the dipole-vortex case, here intervortex lengths will reach their extremum values at the critical point simultaneously. This is due to the reduction of the system down into an effective three vortex interaction. Our results at the critical lengths of the vortex configuration are presented in table~\ref{tab:4_four_vortex_critical_lengths}. Separations for vortices with the fourth vortex are dependent on the vortex separations between other vortices on account of the additional geometric constraints and can easily be recovered from the aforementioned conditions. As such they are not presented here.

\begin{figure}[htp!]
\begin{center}
	\includegraphics[width=0.7\textwidth]{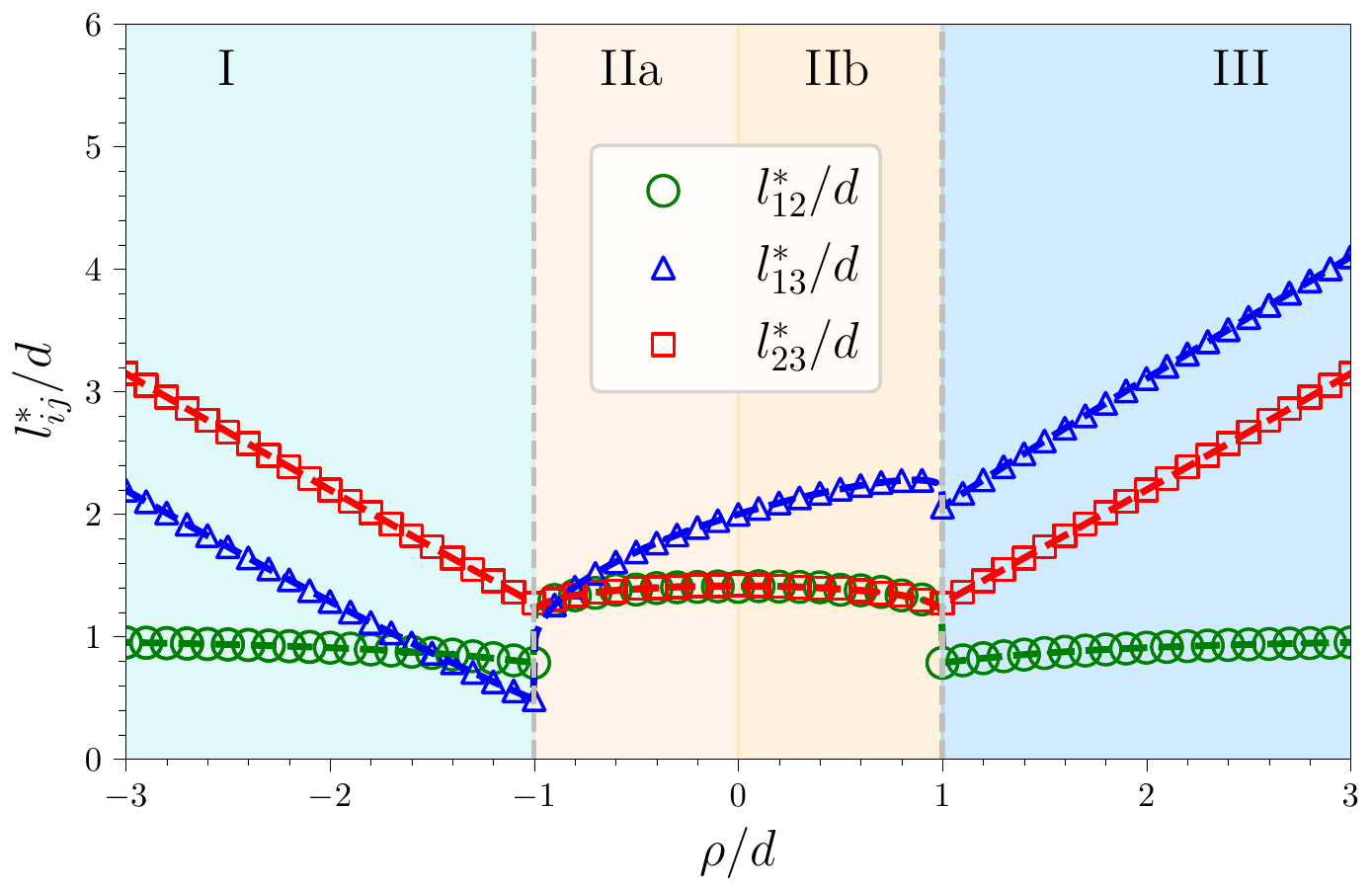}
	\caption{Critical vortex separation lengths in the integrable dipole-dipole interaction.  Theoretical results are plotted as dashed curves given by the formulas in table~\ref{tab:4_four_vortex_critical_lengths} with numerical results as colored symbols. The background is shaded and labeled according to the regions in table \ref{tab:3_four_vortex_regions}. Boundaries between regions represented as dashed gray lines indicate the change from exchange and direct scattering process. \label{fig:7_four_vortex_critical_lengths}}
	\end{center}
\end{figure}

\begin{table}[htp!]
\begin{center}
	\caption{Critical vortex separation lengths of the reduced three vortex system in the integrable dipole-dipole collision categorized by the regions defined in table~\ref{tab:3_four_vortex_regions}.\label{tab:4_four_vortex_critical_lengths}}
\setlength{\tabcolsep}{5pt}
\renewcommand{\arraystretch}{1.25}
\begin{tabular}{c|c|c|c}
	Region & $l^*_{12}$ &  $l^*_{13}$ & $l^*_{23}$\\
	\hline 
	 I &$d\sqrt{  -\frac{\rho}{d}\sqrt{1+\frac{1}{4}\frac{\rho^2}{d^2}} - \frac{1}{2}\frac{\rho^2}{d^2}}$ 
	 & $d\sqrt{-\frac{\rho}{d}\left[ \sqrt{4+\frac{\rho^2}{d^2}}-2\right]}$ & 
    $d\sqrt{  -\frac{\rho}{d}\sqrt{1+\frac{1}{4}\frac{\rho^2}{d^2}} + \frac{1}{2}\frac{\rho^2}{d^2}}$ \\

	IIa & $d\sqrt[4]{2+2\sqrt{1-\frac{\rho^2}{d^2}}}$ & 
	$d\sqrt{2\frac{\rho}{d} + 2\sqrt{2}\sqrt{1+\sqrt{1- \frac{\rho^2}{d^2}}}}$ &
	$d\sqrt[4]{2+2\sqrt{1-\frac{\rho^2}{d^2}}}$\\

	IIb & $d\sqrt[4]{2+2\sqrt{1-\frac{\rho^2}{d^2}}}$ & 
	$d\sqrt{2\frac{\rho}{d} + 2\sqrt{2}\sqrt{1+\sqrt{1- \frac{\rho^2}{d^2}}}}$ &
	$d\sqrt[4]{2+2\sqrt{1-\frac{\rho^2}{d^2}}}$\\

	 III &$d\sqrt{  \frac{\rho}{d}\sqrt{1+\frac{1}{4}\frac{\rho^2}{d^2}} - \frac{1}{2}\frac{\rho^2}{d^2}}$ 
	 & $d\sqrt{\frac{\rho}{d}\left[ \sqrt{4+\frac{\rho^2}{d^2}}+2\right]}$ & 
    $d\sqrt{  \frac{\rho}{d}\sqrt{1+\frac{1}{4}\frac{\rho^2}{d^2}} + \frac{1}{2}\frac{\rho^2}{d^2}}$ \\	
 \end{tabular}

\end{center}
\end{table}
In figure~\ref{fig:7_four_vortex_critical_lengths} we plot the theoretical predictions of the critical vortex separation lengths presented in table~\ref{tab:4_four_vortex_critical_lengths} against numerical data obtained from our simulations. We observe complete agreement between theory and numerical data across all regions of parameter space. We can ascertain that the initial dipole composed of vortices $1$ and $2$ shrinks experiencing a direct scattering process when the parameters are in region I and III with the closest critical distance being attained when the impact parameter $\rho/d=-1$. Interestingly, the minimal dipole size across the parameter space is larger than what can be produced in the dipole-vortex interaction at a distance of $l_{12} = 0.79d$ as opposed to $d/2$. Interestingly, at the same value of the impact parameter  $\rho/d=-1$ the two positive point vortices $1$ and $3$ approach even closer at a distance of $l_{13}=0.49d$ at a point where the vortex configuration is in a complicated and tight four vortex structure. In regions IIa and IIb, we observe exchange scattering ($l_{12}=l_{23}$) and where the two dipoles come together and exchange vortices creating two new sets of dipoles that then propagate away with both new dipoles relaxing back to their initial sizes of $d$ as ensured by the conservation of energy.

Indeed, the long-time dynamics of the system after scattering is that of two oppositely propagating dipoles of equal size $d$ due to the geometric constraint of the parallelogram (which we have confirmed numerically). In general, this behavior does not occur in the non-integrable dipole-dipole collision that we will investigate in the next subsection.

\begin{figure}[htp!]
	\begin{center}
		\includegraphics[width = \textwidth]{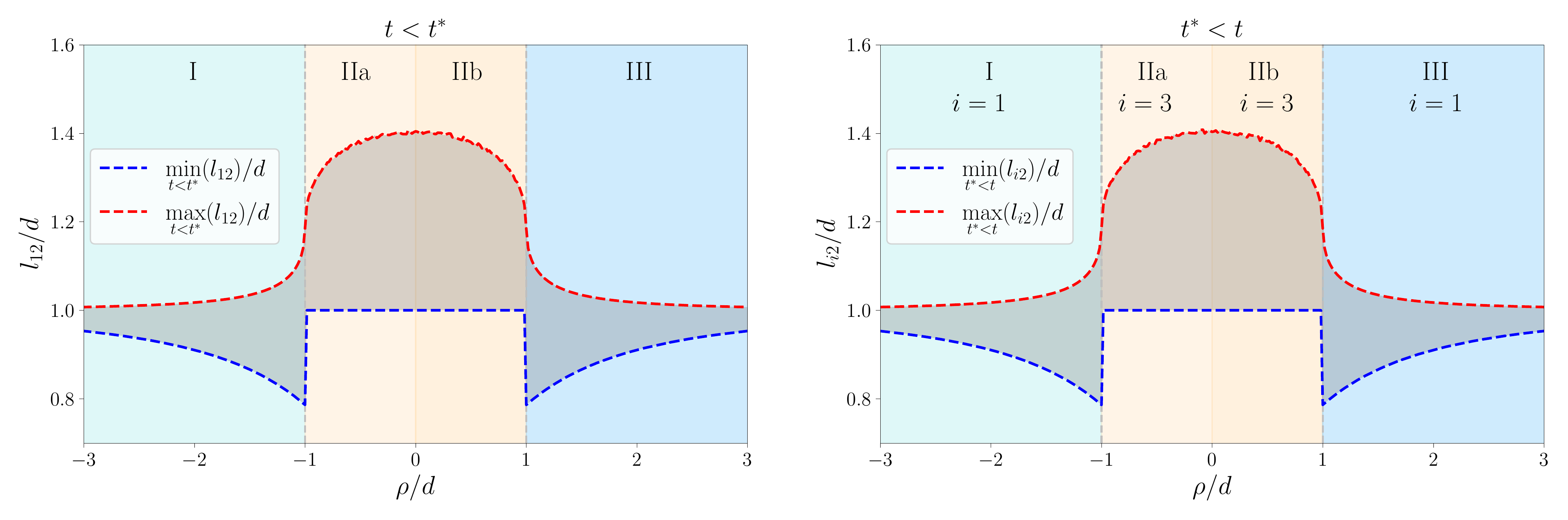}
		\caption{Numerical results of the minimum and maximum size of the propagating vortex dipole pre- and post-interaction defined as the moment $t^*$ that the critical point is reached. Post-interaction we track the vortex dipole that includes the negative point vortex $2$. \label{fig:8_four_vortex_integrable_dipole_size}}
	\end{center}
\end{figure}

In figure~\ref{fig:8_four_vortex_integrable_dipole_size} we show the minimum and maximum distances of the vortex dipole involving point vortex $2$. We observe qualitatively similar behavior to the integrable dipole-vortex collision with the dipole shrinking during direct scattering, but growing during exchange scattering. The pre- and post-interaction symmetry arises due to the critical points occurring at the moment of collinearity or exchange (see in the next subsection that this is not necessarily the case in general).

\subsection{Non-Integrable Dipole-Dipole Scattering}
\label{sec:non-int-4vortex}

\begin{figure}[htp!]
\begin{center}
	\includegraphics[width=0.7\textwidth]{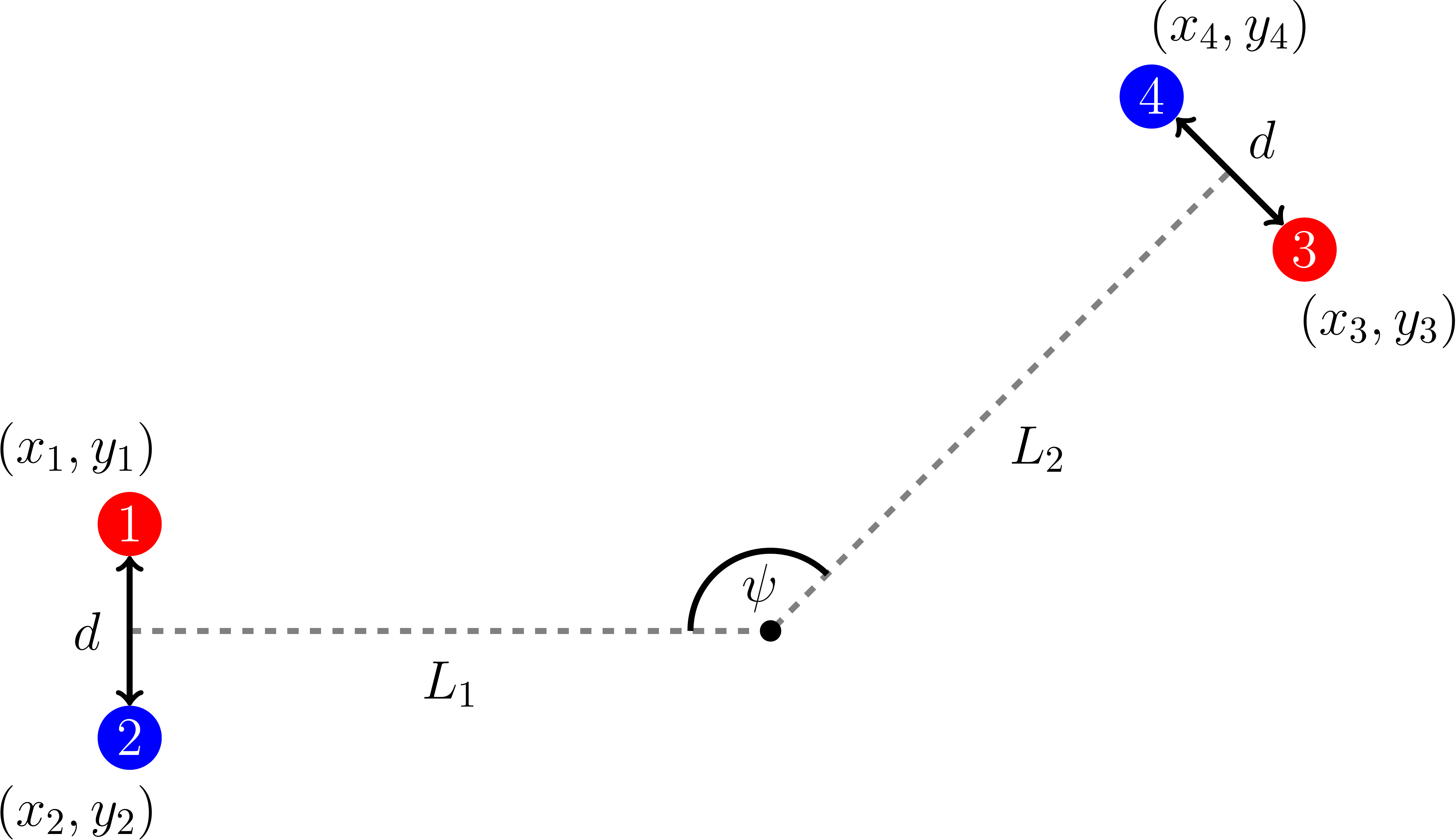}
	\caption{Setup of the non-integrable four vortex dipole-dipole collision defined by two parameters, the ratio of the dipole separations from the origin  $\delta_L=L_1/L_2$, and the angle of incidence $\psi$. The four vortices are arranged such that two dipoles of size $d$  with both orientated such that their trajectories intersect at the origin. \label{fig:9_four_vortex_chaotic_setup}}
\end{center}
\end{figure}

\noindent The general non-integrable equally-sized, equal circulation dipole-dipole setup is displayed in figure~\ref{fig:9_four_vortex_chaotic_setup}. Here, two dipoles of size $d$ are situated away from the origin by distance $L_1$ and $L_2$ respectively and orientated with respect to each other by an angle $\psi$ around the origin. 

The general initial configuration is fully given by the angle $\psi \in [0,2\pi)$ and the lengths $L_1, L_2 \in \left(0,\infty\right)$. As we are interested in the limit of $L_1, L_2 \to \infty$, we define the ratio $\delta_L = L_1/L_2$, as the limit is taken. Moreover, due to the symmetry of the problem, we find that the transformation of $\psi\to 2\pi-\psi$ leads to the original configuration if we relabel vortices $1\leftrightarrow 3$ and $2\leftrightarrow 4$ respectively.  As this configuration is non-integrable for $\psi\neq \pi$~\cite{eckhardt_irregular_1988,eckhardt_integrable_1988} we are unable to use analytical methods as in subsection~\ref{sec:int-4vortex}. Therefore our investigation will be solely numerical in nature. In~\cite{eckhardt_irregular_1988} Eckhardt performed a brief numerical study highlighting the rich dynamics of two interacting dipoles in the non-integrable case for both direct and exchange scattering scenarios. Our goal is to extend this numerical study and examine other aspects of the dipole dynamics, such as the dipole scattering angle and the evolution of the dipole size during and after interaction.

The initial vortex configuration of figure~\ref{fig:9_four_vortex_chaotic_setup} gives the following conserved quantities in the limit $L_1,L_2\to \infty$: $
H \stackrel[L\to\infty]{}{\rightarrow} 
\frac{\kappa^2}{2\pi}\ln\left(d^2\right)
$ which leads to the usual value of the Hamiltonian for two infinitely separated dipoles of size $d$, and  linear momentum ${\bf P}=\left\{ -\kappa d \sin(\psi), -\kappa d \left[1+ \cos(\psi)\right] \right\}$, and angular momentum $M = 0$, leading to $R = -2\kappa^2d^2\left[1 +\cos(\psi) \right]$. The values of the linear momentum ${\bf P}$ and subsequently $R$ are related to the orientation of the right-hand side dipole in the upper half plane. A simple translation of the angle $\psi$ can ensure that the linear momentum vanishes and the integrable four vortex case with $C=0$  is recovered.

We perform a series of numerical simulations of the point vortex model with initial conditions as presented in figure~\ref{fig:9_four_vortex_chaotic_setup} for several values in the parameter space of $\psi$ and $\delta_L$ with $L_1,L_2 \gg d$. All numerical simulations are performed in the non-integrable case such that the relative error of the Hamiltonian is conserved within $10^{-10}$.

\begin{figure}[htp!]
\begin{center}
	\includegraphics[width=0.7\textwidth]{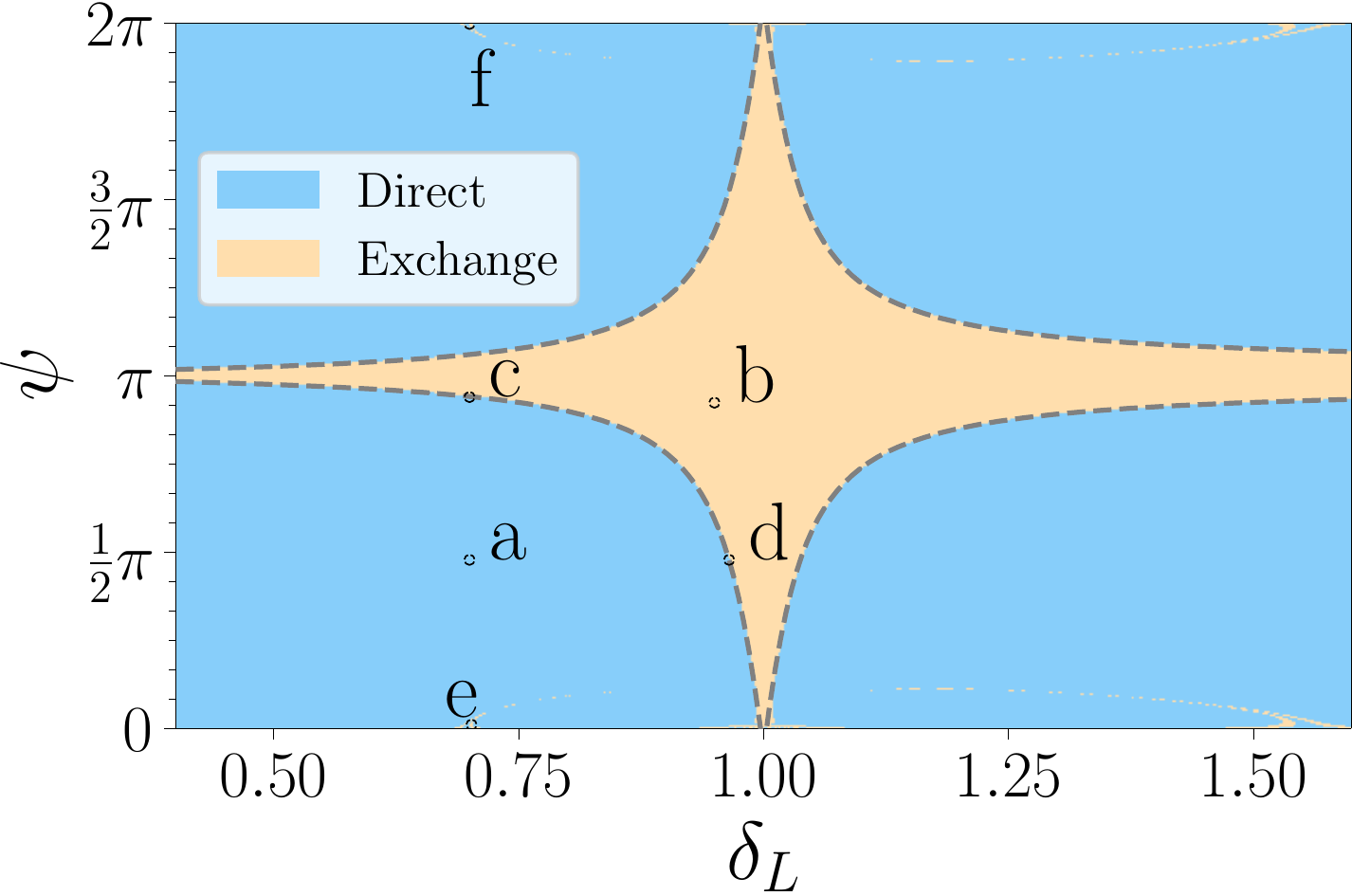}
	\caption{Colormap of the type of interaction observed in the non-integrable dipole-dipole collision across the parameter space. Light blue regions indicate parameter regions of direct scattering while regions of yellow correspond to exchange scattering where the initial dipoles exchange vortices during the interaction. Six particular regions of interest have been highlighted and labeled (a-f). The boundary between main direct and exchange scattering regions are marked by gray dashed curves.\label{fig:10_four_vortex_chaotic_regions}}
\end{center}
\end{figure}

\begin{figure}[htp!]
	\begin{center}
		\includegraphics[width=\textwidth]{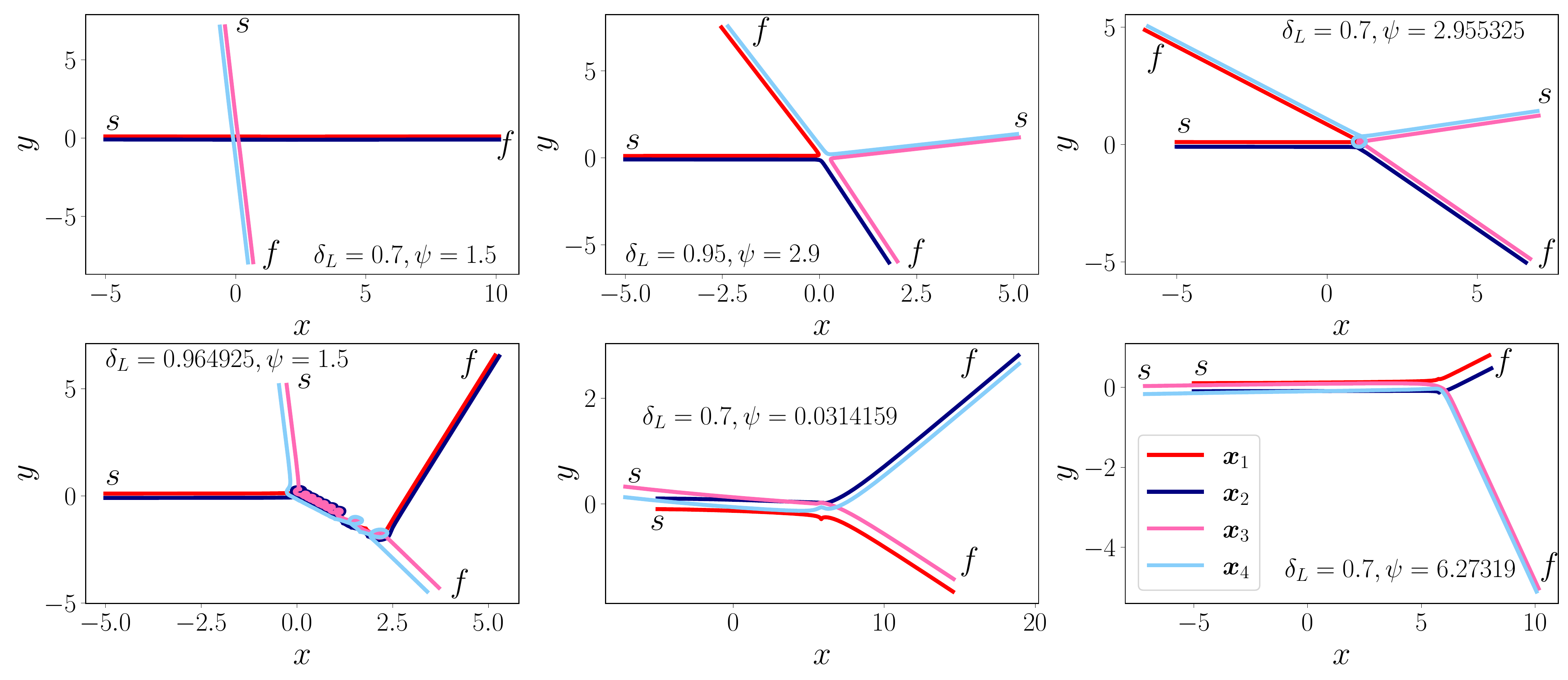}
		\caption{Vortex trajectories from six parameter sets (top row, left to right) (a) $\psi =1.5, L=0.7$, (b) $\psi =2.9, L=0.95$, (c) $\psi =2.955325, L=0.7$, (bottom row, left to right) (d) $\psi =1.5, L=0.964925$, (e) $\psi =0.0314159, L=0.7$, and (f) $\psi-6.27317, L=0.7$. Labels $s$ and $f$ indicate the start and finishing points of the simulation. The dark red and blue curves correspond to vortices $1$ and $2$, with the pink and light blue curves vortices $3$ and $4$. \label{fig:11_four_vortex_chaotic_trajectories}}
	\end{center}
\end{figure}

We begin by characterizing regions of direct and exchange scattering by monitoring dipole composition in the long time limit. A map of the interaction types observed is presented in figure~\ref{fig:10_four_vortex_chaotic_regions}. The majority of exchange scattering occurs in a star shaped region (yellow) centered around $\psi=\pi$ and $\delta_L=1$. This is to be expected as for values of $ \delta_L\approx 1$  the two dipoles will be orientated in such a way that a collision with dipoles becoming very close will occur for any angle $\psi$. Moreover, if $\psi\approx \pi$ then we are close to the integrable dipole-dipole collision with small impact parameter which indicates that an exchange scatter process will be likely.  We observed in both the integrable dipole-dipole collision as well as the dipole-vortex collision that exchange scattering typically occurs when dipoles undergo extreme collisions with another vortex. Outside the internal star region (light blue), we observe that the dipole propagation paths are such that an extreme collision is unlikely and a direct scattering process occurs with the two dipoles interacting from a distance leading to a slight deflection in their prorogation path without destabilizing the dipole structure.  Interestingly, there are regions of exchange scattering close to $\psi =0$ and $\psi=2\pi$ that swoop out towards the center. We have highlighted six specific interactions (a-f) in our parameter space in which we display the observed vortex interactions. In figure~\ref{fig:11_four_vortex_chaotic_trajectories} we plot the trajectories of the four vortices during interaction at these specific parameter points. Figure~\ref{fig:11_four_vortex_chaotic_trajectories}(a) shows the two dipoles missing each other with very little deflection due to the small value of $\delta_L\approx 0.7$. Figures~\ref{fig:11_four_vortex_chaotic_trajectories}(b-c) are two types of exchange scattering where a pair of vortices are exchanged between the dipoles. Notice how the interaction is different in figure~\ref{fig:11_four_vortex_chaotic_trajectories}(c) (which is situated close to the boundary between direct and exchange scattering) displaying a complicated rotation between the two dipoles close to the moment of exchange. Interestingly, if we take another point close to the direct and exchange boundary figure~\ref{fig:11_four_vortex_chaotic_trajectories}(d) we observe similar but more prolonged four vortex dynamics composed of a rotational dance before the two dipoles propagate away. It is this swirling motion that leads to extremely large values of scattering angle for the interaction. Moreover, notice that the relative sizes of the two dipole have changed after undergoing the direct scattering which was something banned in the integrable cases. In figures~\ref{fig:11_four_vortex_chaotic_trajectories}(e) and (f) we take two sets parameters close to the $\psi=0$ and $\psi=2\pi$ boundaries. This means that both dipoles in both realizations are close to propagating along the same axis. We observe in figure~\ref{fig:10_four_vortex_chaotic_regions} that both (e) and (f) border the swooping exchange regions meaning that the type of orientation is sensitive to the scattering process. We see an indication of this in figures~\ref{fig:11_four_vortex_chaotic_trajectories}(e) and (f) where the former is exchange scattering and the latter is direct scattering. In both cases (e) and (f) a dipole begins propagating behind the initial dipole, at a very small incidence angle, whereupon as the system evolves the behind dipole eventually catches up to the dipole in front. If parameters are such that they reside on one of the swooping ``shark-fin" exchange scattering curves, we then observe exchange scattering as shown in (e). Otherwise for parameters off this curve we see the behind dipole deflecting the front dipole off the initial trajectory with no exchange of vortices, as seen in (f), with both dipoles propagating off in a direct scattering process. We observe, in both cases (e) and (f), a change in the final dipole sizes (one increasing, the other decreasing in size). The relative change of the dipole sizes is controlled by the conservation of the Hamiltonian as the final state when the dipoles are infinitely separated must still  lead to $H= (\kappa^2/2\pi) \ln(d^2)$ (we will discuss more on this later).

There are also now two different boundary interactions that occur at the border between exchange and direct scattering. The first, indicated in figure~\ref{fig:11_four_vortex_chaotic_trajectories}(c), corresponds to Havelock's alternating vortex rings~\cite{havelock_stability_1931} that has already been encountered in the integrable case (section~\ref{sec:int-4vortex}). The type of interaction observed in figure~\ref{fig:11_four_vortex_chaotic_trajectories}(d) however has not been encountered before: the motion consists of vortices $1$, $2$ and $3$ forming a rotating structure made up of a two positive and one negative vortex giving a quasi-stable three vortex structure with a total circulation of $\kappa$, which then essentially forms a vortex/anti-vortex pair with the remaining vortex $4$ until the structure eventually destabilizes and returns to two coherent dipoles of differing sizes.

\begin{figure}[htp!]
\begin{center}
	\includegraphics[width=0.8\textwidth]{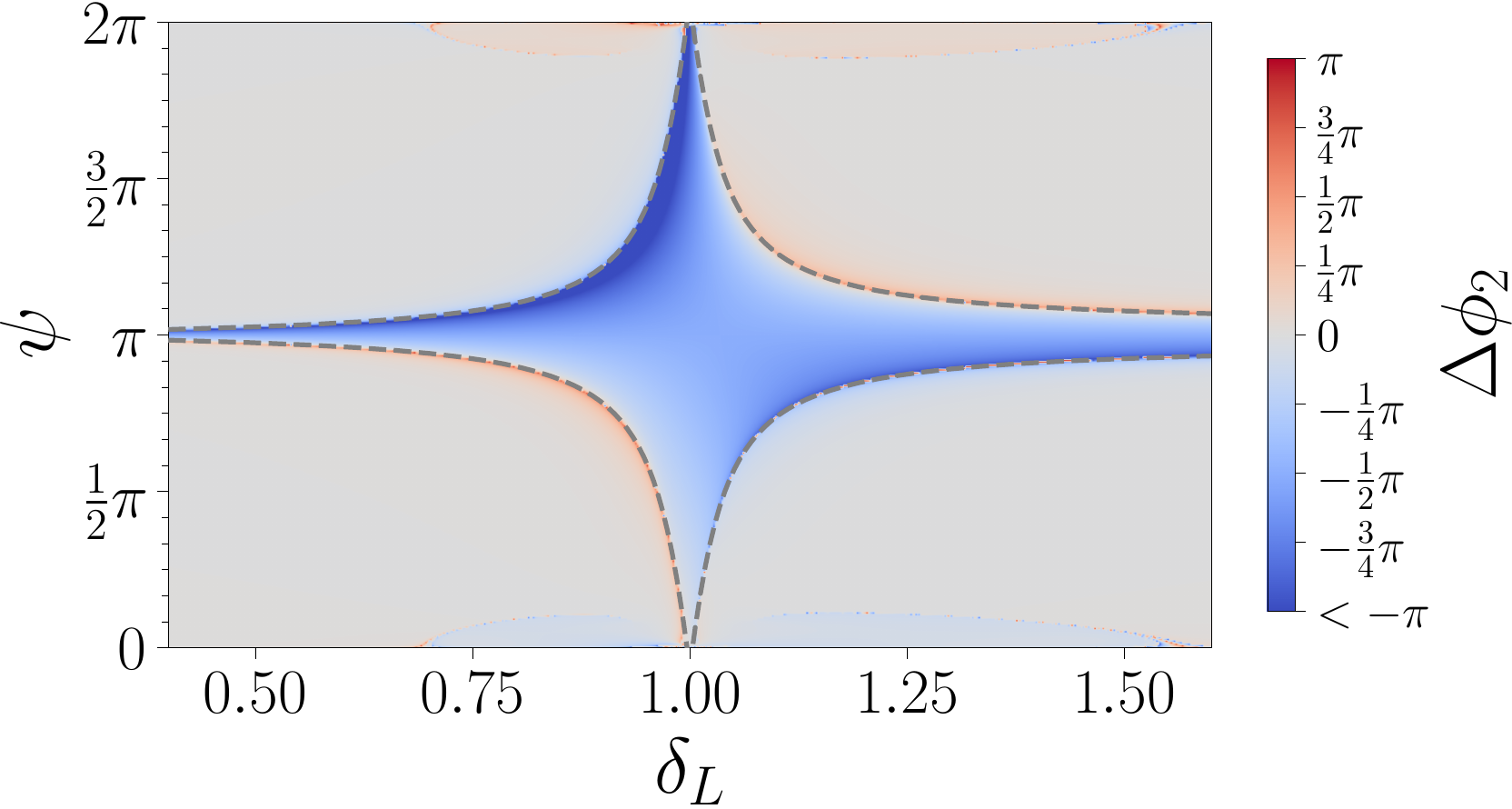}
	\caption{Heatmap of the scattering angle of the dipole-dipole interaction over the parameter space. Results are visualized as a colorbar centered at zero, which is taken as the horizontal axis. Region boundaries between direct/exchange scattering are marked by gray dashed curves.\label{fig:12_four_vortex_nonintegrable_angles}}
\end{center}
\end{figure}

 As with the previous numerical simulations, we track the direction of propagation of vortex $2$ and determine the corresponding deflection angle $\Delta \phi_2 = \lim_{t\to\infty} \phi_2(t) - \lim_{t\to-\infty} \phi_2(t)$ measured only after sufficient time has elapsed post-interaction to ensure that the vortex dipole is isolated and is propagating only via self-interaction. The scattering angle results are presented in figure~\ref{fig:12_four_vortex_nonintegrable_angles}. Notice that in the direct scattering regions, there is very little deflection of the dipole propagation direction when compared to the  dipole-vortex collision. This is a consequence of interaction strength (velocity) of a dipole decaying as $\propto 1/r^2$, while for a single vortex the decay is $\propto 1/r$. This means that the far-field interaction between two dipoles is weaker than what is experienced in the dipole-vortex scattering process and so significant scattering only occur when the two dipoles are close or directly impacting. Consequently, we observe significant scattering mainly in or close to the central exchange scattering region. The inner exchange scattering region is predominantly blue meaning that after an exchange of vortices, the vortex dipole containing vortex $2$ exhibits a negative angle deflection, i.e. a clockwise deflection. There is a small band of positive angle deflection along the direct-exchange border region as indicted by the realization displayed in figure~\ref{fig:11_four_vortex_chaotic_trajectories}(d). A small band of red is observed in the upper left region of the inner exchange region. Notice the sharp transition from red to dark blue which is due to the $2\pi$ winding of our deflection angle.

Overall,  the scattering between two equal strength dipoles is much more regular than what is observed for dipoles of different strengths. Indeed, the chaotic scattering studied by Aref and Pomphrey~\cite{aref_integrable_1980} is not observed here. With that being said, there are significant jumps in the scattering angle $\Delta\phi_2$ at particular points in the parameter space, most notably close to the boundary of the inner direct-exchange region in the upper left-hand quadrant. The dynamics of the interaction in these cases involve a longer four vortex interaction that includes repeated rotation of the vortex system, e.g. the cases presented in figure~\ref{fig:11_four_vortex_chaotic_trajectories}(c) and (d), and is similar in style to the integrable four vortex case where these boundaries regions correspond to Havelock's double alternating rings~\cite{havelock_stability_1931}.

Ultimately, we observe that in the dipole-dipole collision that the direct scattering process is less effective compared to the dipole-vortex collision due to the reduced far-field strength of a dipole compared to an isolated vortex.  However, in the dipole-dipole case, the exchange scattering process is significantly more disruptive to the initial state showing a variety of complex four vortex interactions.

\begin{figure}[htp!]
	\begin{center}
	\includegraphics[width=\textwidth]{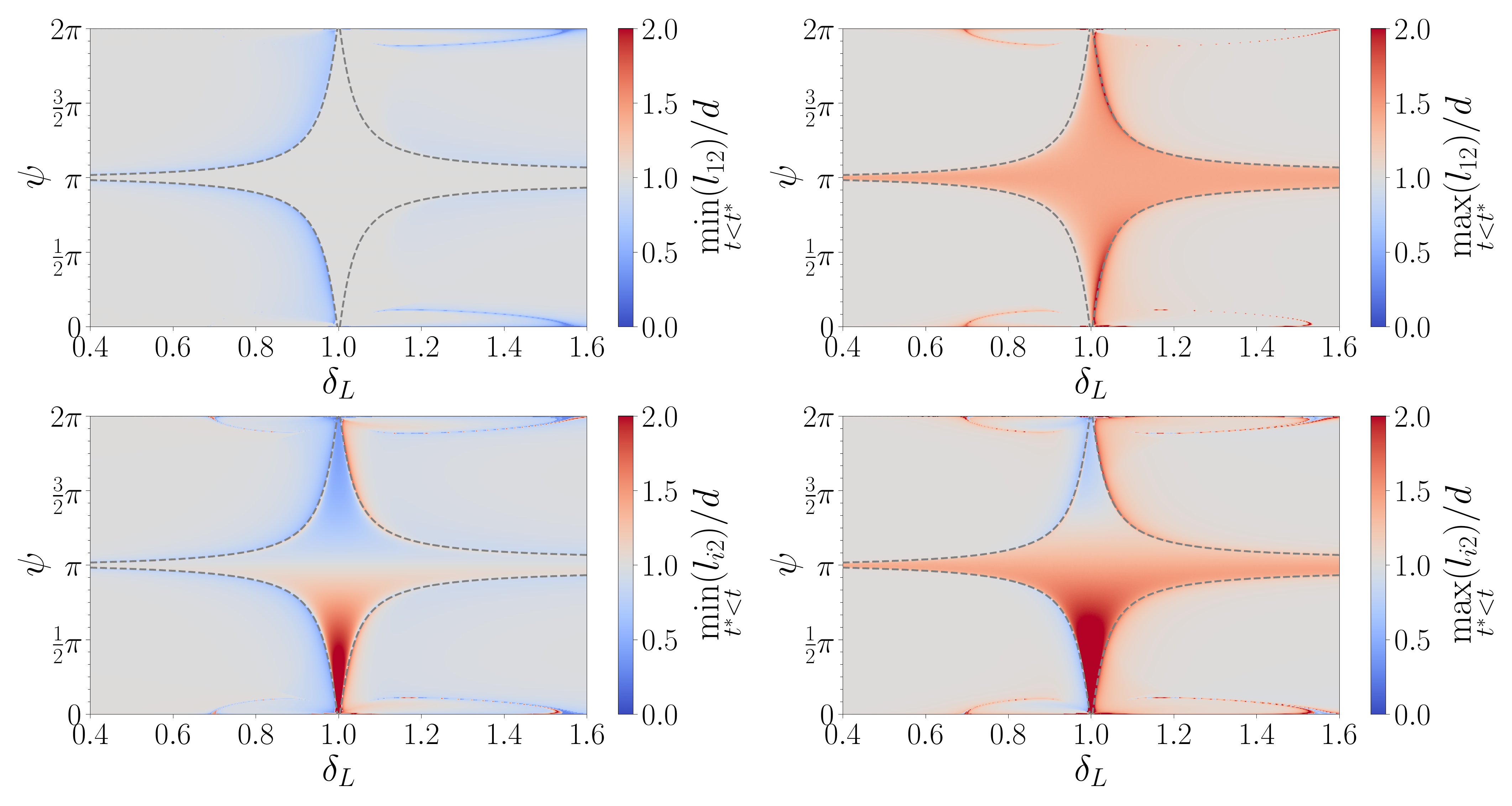}
	\caption{Minimal (left) and maximal (right) dipole lengths pre- (top) and post-interaction (bottom) in the non-integrable dipole-dipole collision.
	\label{fig:13_four_vortex_nonintegrable_dipole_size}}
	\end{center}
\end{figure}

In figure~\ref{fig:13_four_vortex_nonintegrable_dipole_size} we plot the minimum and maximum dipole size pre- and post-interaction defined by the time $t^*$ which is the time for which the length $l_{13}$ reaches a critical point, with pre-interaction defined as the phase of motion $t<t^*$ and $t>t^*$ defined as the post-interaction phase. We observe that the initial dipole size $l_{12}$ can reduce in magnitude within the pre-interaction stage during a direct scattering process, in particular with parameters close to the direct-exchange boundary. In the exchange region, we observe that the minimal distance remains $l_{12}/d=1$ which indicate that the dipole grows, as indicated by the red coloring in the maximal distance pre-interaction (figure~\ref{fig:13_four_vortex_nonintegrable_dipole_size} top right). Interestingly, post-exchange interaction we have cases of the newly formed dipole either shrinking ($\psi \gtrsim \pi$) or enlarging ($\psi \lesssim \pi$) depending on the initial condition parameters. 

\begin{figure}[htp!]
\begin{center}
	\includegraphics[width=\textwidth]{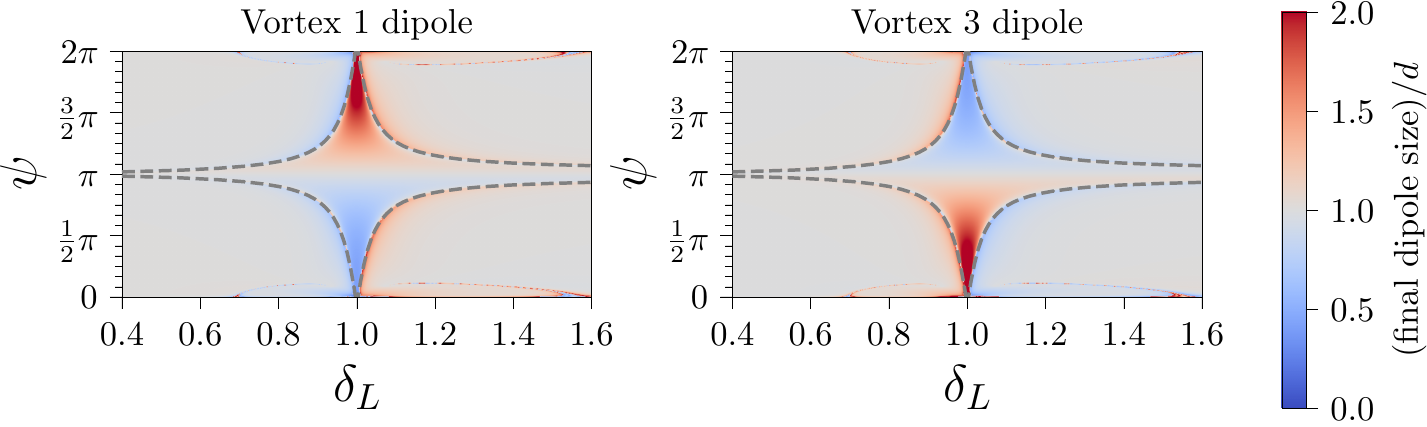}
\caption{Final dipole separations post-interaction of each vortex dipole (identified by containing vortex $1$ or $3$). Each heatmap shows the normalized dipole size by the initial dipole separation $d$. The boundary between the main direct-exchange region is shown by the gray dashed curve.\label{fig:14_four_vortex_nonintegrable_dipole}}
\end{center}
\end{figure}

in figure~\ref{fig:14_four_vortex_nonintegrable_dipole} we plot the final dipole sizes after interaction. Due to the conservation of $H=(\kappa^2/2\pi) \ln(d^2)$, the final state is always of two dipoles of sizes $d_1$ and $d_2$ such that $d_1d_2=d^2$. This because each dipole contributes $(\kappa^2/2\pi) \ln(d_i)$, $i=1,2$ to the system energy.   This can be observed in figure~\ref{fig:14_four_vortex_nonintegrable_dipole} from the symmetry of the two dipole distances. We confirmed that the product of the normalized distances equals $1$ for all values of the parameter $\psi$ and $\delta_L$. Moreover, by averaging over the parameter range, we also confirm that there is no preference in the creation of larger or small dipoles, with the average distance being equal to $d$. For the non-integrable dipole-dipole collision,  the closest a dipole can form is of size $0.016d$ at parameters $\psi=6.28319$ and $\delta_L=1.00188$, which is an extreme configuration of two closely chasing vortex dipole aligned along the same axis of propagation. 

In summary, the non-integrable dipole-dipole collision is the simplest collision where it is possible for a dipole to become significantly smaller or larger than the initial dipole after interaction. In the non-integrable dipole-dipole collision energy conservation is maintained by the relative (opposing) change of the second dipole size. We also see the most complex interactions occurring near the direct-exchange boundary. This is no more apparent than in the non-integrable dipole-dipole collision where we have shown examples of a wide array of complex  interactions involving the formation of quasi-stable three vortex structures and the like. Furthermore, we have shown that direct scattering of a dipole is stronger when with an isolated vortex due to the weaker decay ($\propto 1/r$) of an isolated vortex compared to a secondary dipole $\propto 1/r^2$.

\section{The Dipole-Cluster Collisions}
\label{sec:dipole-cluster}

\noindent In this section we investigate the interaction of a dipole with an $m$-sized cluster, that is, a coherent vortex structure consisting of $m$ same-sign vortices. To be more specific, we study symmetric clusters with $m=2,3,4$ identical vortices labeled $C_m$. Of particular interest here are questions of cluster stability, approximation by comparison with the dipole-vortex interaction, new types of interaction that may be possible in such systems, and the distribution of dipole sizes for such systems. These problems can be considered the most complex of the basic vortex structure interactions due to the additional degrees of freedom, and will be more relevant to large vortex configurations that more appropriately resemble turbulent flow.

\subsection{Dipole-Cluster Scattering}

\begin{figure}[htp!]
\begin{center}
\includegraphics[width = 0.9\textwidth]{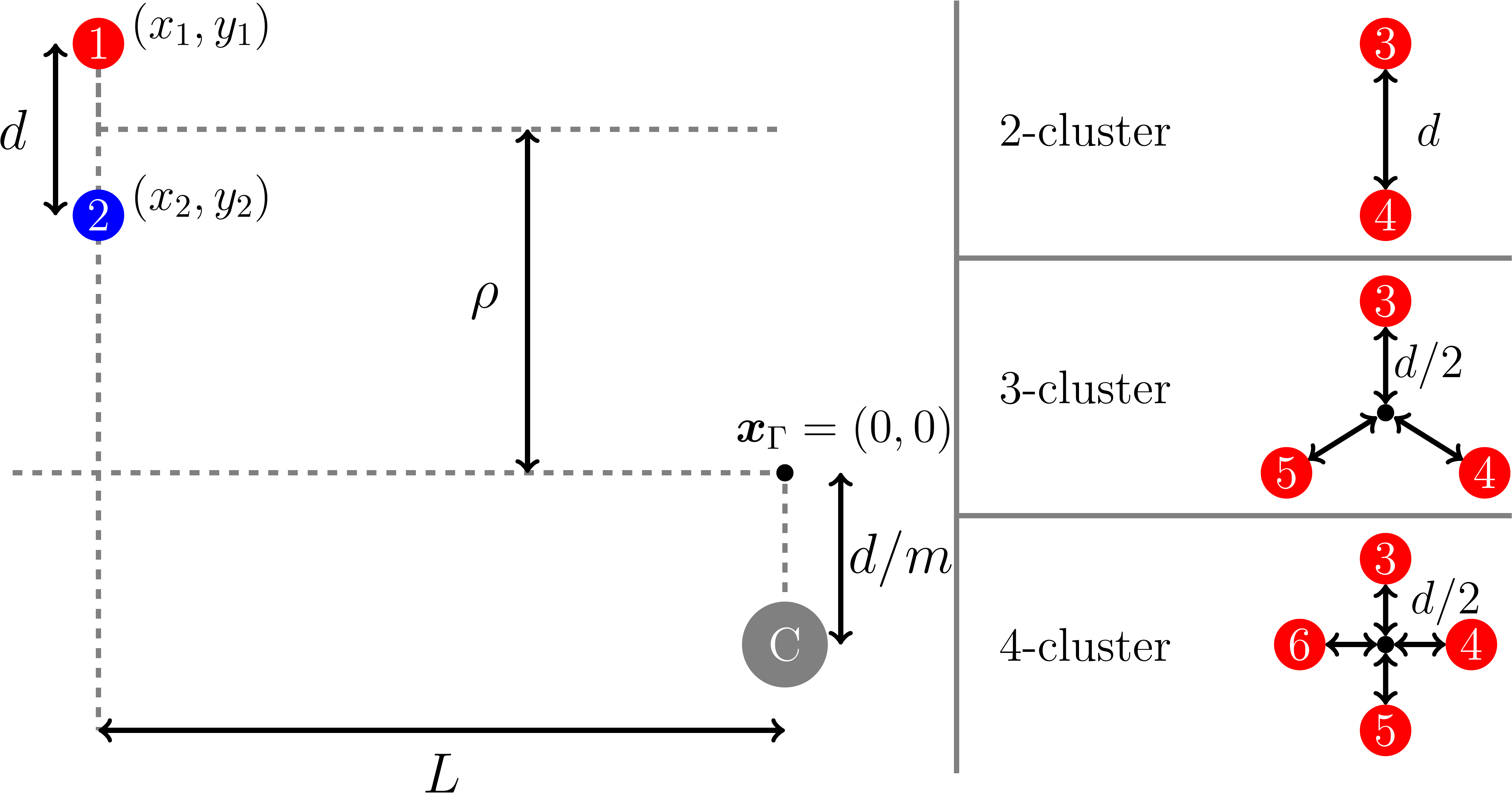}
\caption{Initial setup of the dipole-cluster collision. Vortices $1$ and $2$ form a dipole of size $d$ situated a distance $L$ from the cluster $C$. The quantity $\rho$ acts as an impact parameter, measuring the vertical distance from the center of the vortex dipole to the center of circulation. The cluster structures $C_2, C_3, C_4$ are presented on the right.  \label{fig:15_cluster_setup}}
\end{center}
\end{figure}

\noindent A schematic of the initial configuration used in the dipole-cluster collision is presented in figure~\ref{fig:15_cluster_setup}. The idealized and symmetric structures of the vortex cluster $C_m$ for values  $m=1,2,3$ are also presented. The system is initialized such that a dipole of size $d$ is situated a large distance $L$ from the cluster, propagating towards the self-induced rotating $m$-vortex cluster. The impact parameter $\rho$ is defined as the vertical distance from the center of the cluster to the dipole. Each $m$-vortex cluster is configured as a regular convex $m$-sided polygon, such that the counter-clockwise rotation of the cluster circumscribes a circle of diameter $d$. Consequently, we introduce an additional variable $\xi$ that represents the rotational phase of the cluster, with $\xi=0$ taken as the cluster oriented as displayed in figure~\ref{fig:15_cluster_setup}, and hence, each cluster will have rotational symmetry isomorphic to the cyclic group of order $m$. Simulations are performed in the cluster cases such that the relative error of the Hamiltonian is conserved to at least $10^{-10}$.

\begin{figure}[htp!]
	\begin{center}
    \includegraphics[width=\textwidth]{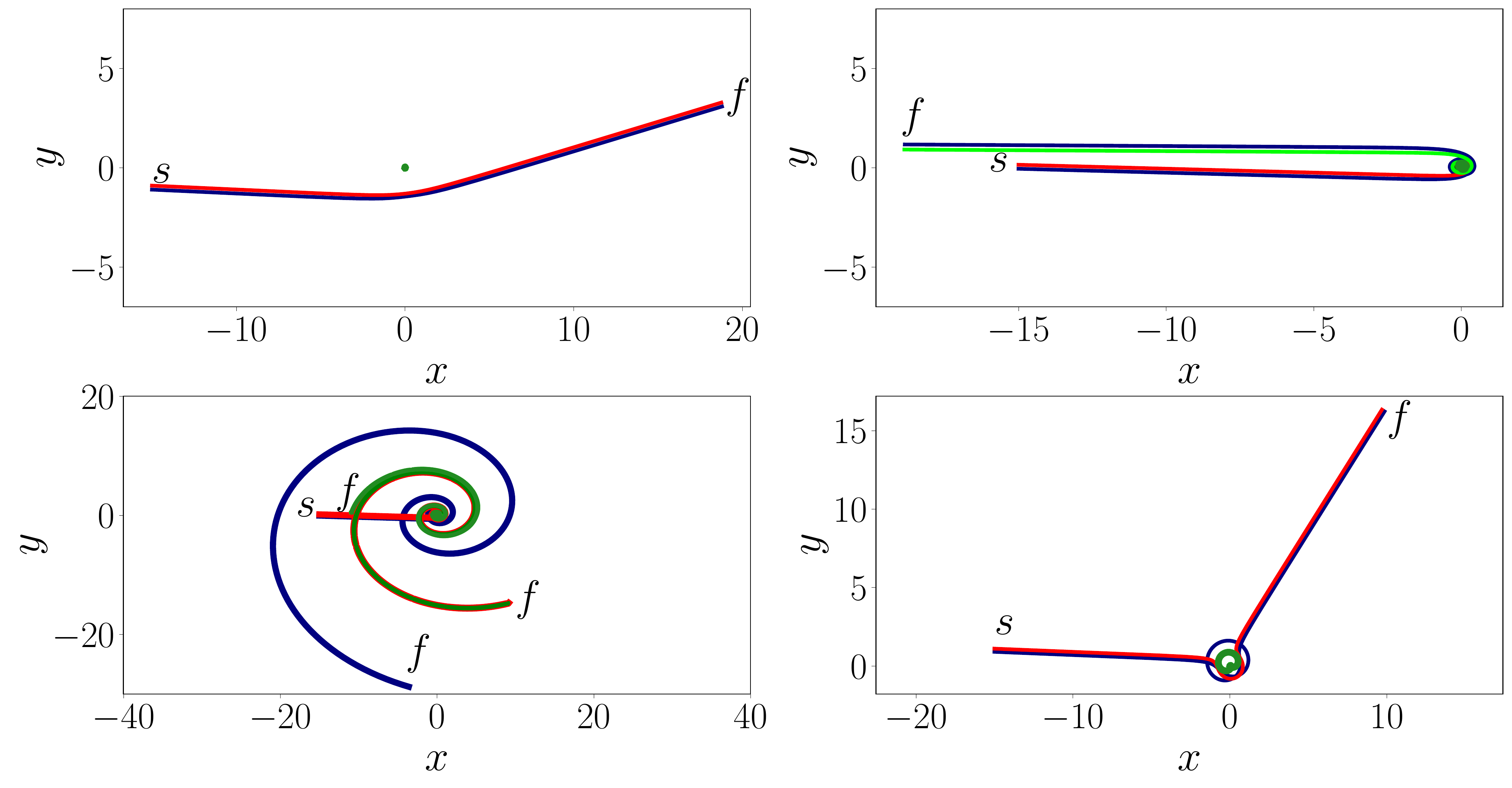}
	\caption{Sample trajectories of the dipole and $C_3$ interaction for four parameter sets. The dipole is initialized a distance of $L=15$ in each case such that the phase of the cluster is the same for each trajectory, we then have impact parameters as such; top left $\rho/d = -5$, top right $\rho/d = 0.25$, bottom left $\rho/d=0.275$, bottom right $\rho/d = 5$. Dipole vortices 1 and 2 are given as red and blue curves respectively, whilst the remaining $3$ vortices initially forming the cluster are given as green curves. Where not immediately obvious the start and finish points of the simulation are denoted as $s$ and $f$. 
	\label{fig:16_cluster_example_trajectories}}
	\end{center}
\end{figure}

For each simulation of the dipole-cluster collision we perform comparative simulations with the corresponding dipole-$m\kappa$-vortex collision which is a similar set up as shown in figure~\ref{fig:15_cluster_setup}, but with the cluster labeled $C$ replaced by a single positive point vortex of circulation $m\kappa$ for $m=2,3,4$, whose interaction with the dipole should directly correspond to a generalized three vortex dipole-vortex collision similar in behavior to what we analyzed in section~\ref{sec:three_vortex}. We expect that for large values of the impact parameters $\rho$ that the two setups will result in similar characteristics due to the mean-field interactions of the cluster being closely approximated by a single vortex of circulation $m\kappa$. For smaller impact parameters, the analogy will not be so clear. The cluster structure permits ``vortex stretching" in the sense that the cluster vortices are free to rearrange themselves. Sample trajectories of the complex interactions observed in the vortex-cluster simulations of $C_3$ are presented in figure~\ref{fig:16_cluster_example_trajectories}, where we observe the familiar behaviors of direct and exchange scattering in the top row. 

The bottom two trajectories demonstrate more complex interactions not previously possible. The bottom left shows the cluster breaking up into two $C_2$ clusters, rotating in a continual expanding spiral in long-range interaction with the lone anti-vortex. This never-ending spiral expansion is reminiscent of a time-reversed trajectory of a self-similar vortex collapse~\cite{kudela_self-similar_2014}. If the time-reversed form of our trajectory is considered, we would observe three vortex structures spiraling towards the center of vorticity, with this process eventually interrupted as a tight dipole structure is created which then propagates off to infinity. This may be due to the vortex collapse solution being unstable, avoiding a true collapse if slightly perturbed, leading instead to a dramatic change in the distribution of vortex structures.  
In the bottom right we have an example of what we label a pseudo-exchange interaction: where the dipole undergoes a series of vortex exchanges with vortices of the cluster, but ultimately leaves the cluster as a coherent dipole composed of the same initial two vortices. 

Of the four example interactions presented in figure~\ref{fig:16_cluster_example_trajectories}, it is only possible to replicate the direct scattering and pseudo-exchange dynamics in the corresponding dipole-$m\kappa$-vortex collision, as these are the only interactions where the fine internal structure of the cluster does not influence evolution of the system. A complex exchange cannot occur because the third isolated vortex is of stronger circulation than the anti-vortex meaning that a vortex dipole cannot be created. Interestingly, the generalized dipole-$m\kappa$-vortex collision remains an integrable system, giving hope for new analytical results that are applicable for vortex-cluster collisions for large impact parameters. 

\begin{figure}[htp!]
	\begin{center}
		\includegraphics[width = \textwidth]{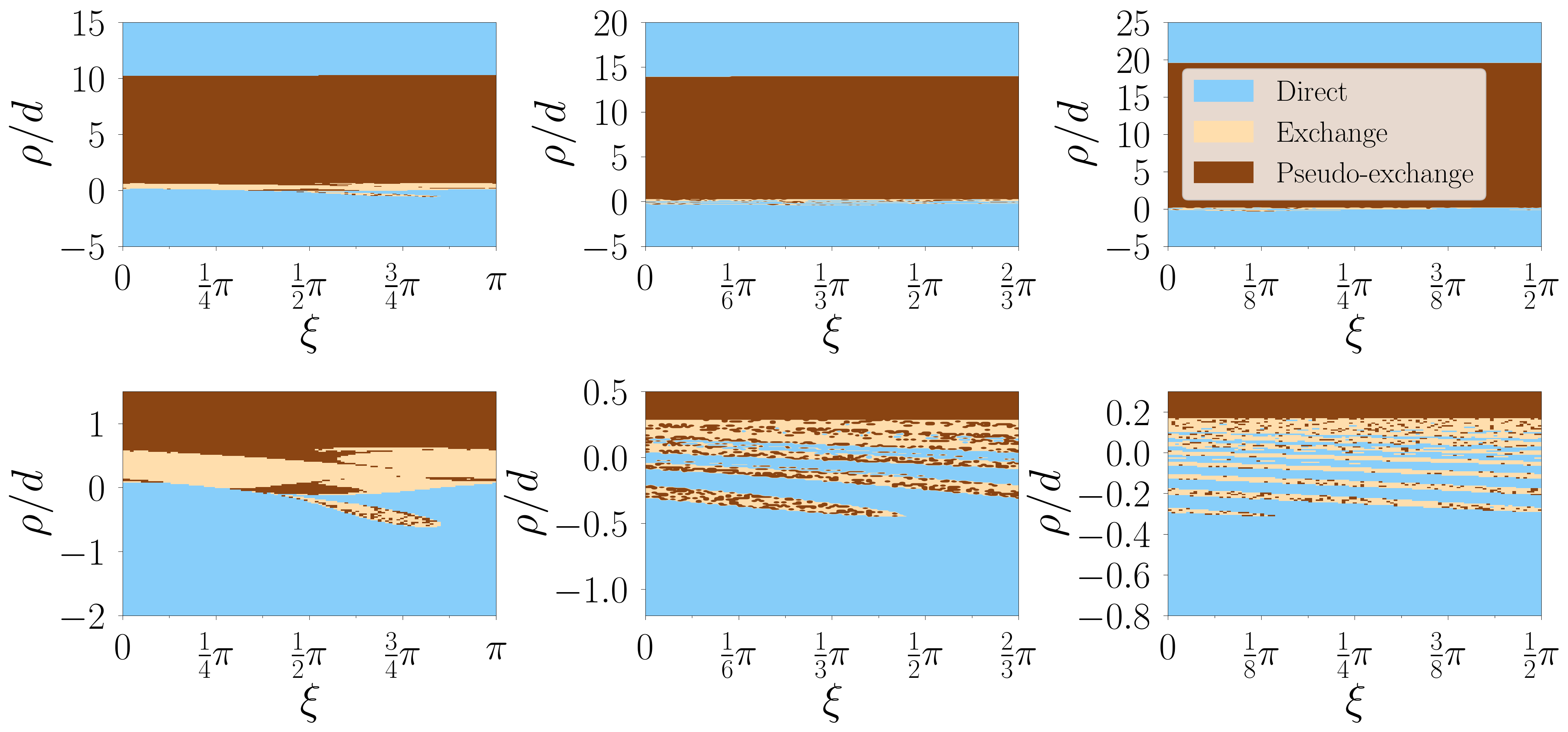}	
		\caption{Colormaps of the dipole-cluster interaction types across the phase $\xi$ and $\rho/d$ for $m=2,3,4$ (left,center,right).  Bottom row are zoomed images of the top row. Direct scattering and exchange scattering are marked in blue and yellow respectively, whilst the new pseudo-exchange interaction type is marked in brown. 
		\label{fig:17_cluster_interaction_regions}}
	\end{center}
\end{figure}

In figure~\ref{fig:17_cluster_interaction_regions} we plot a color-map indicating the types of vortex interactions observed in the dipole-cluster scattering process with respect to the impact parameter $\rho$ and cluster phase $\xi$. For large impact parameters and all values of $m$ we see a clear transition from a direct scattering process to a pseudo-exchange that varies little across the phase $\xi$. This follows from the three vortex interaction as the internal cluster dynamics does not matter significantly when $\rho$ is not close to 0, so the mean field dynamics can be approximated by an $m\kappa$-vortex.  However, there are interesting regions close to $\rho=0$ across all cluster simulations where a narrow band of exchange interactions is spread between the border of direct and pseudo-exchange.

In the bottom row of figure~\ref{fig:17_cluster_interaction_regions} we plot zoomed images of the interaction types around this region along the bottom row. Around $\rho/d=0$ we observe complex band of interaction types composed of exchange and pseudo-exchange scattering interlaced with direct scattering. As the cluster size increases, this band is stretch diagonally across parameter space, wrapping around multiple times in the $C_3$ or $C_4$ cases due to periodicity of $\xi$.

Moreover, as the strength of the cluster, $m$, increases, we observe that this complex region has an extruding tail that becomes thinner meaning that an increasing majority of the parameter space becomes either direct scattering or pseudo-exchange scattering, which by definition implies that the vortices on the incoming dipole and scattered (out-going) dipole are the same. The tail is mottled with patches of pseudo-exchange scattering which appear at a scale comparable to our resolution used in scanning the parameter space. Therefore, there is a possibility that these could comprise of even finer detail. What is particularly surprising is how thin the exchange area is: fractions of the original dipole separation distance $d$. This indicates that it is very unlikely the dipole ends up permanently exchanging vortices with the cluster in these types of interaction; for most of the parameter range we will observe the original dipole propagating away post interaction.

\subsection{Scattering Angles}

\begin{figure}[htp!]
	\begin{center}
		\includegraphics[width=\textwidth]{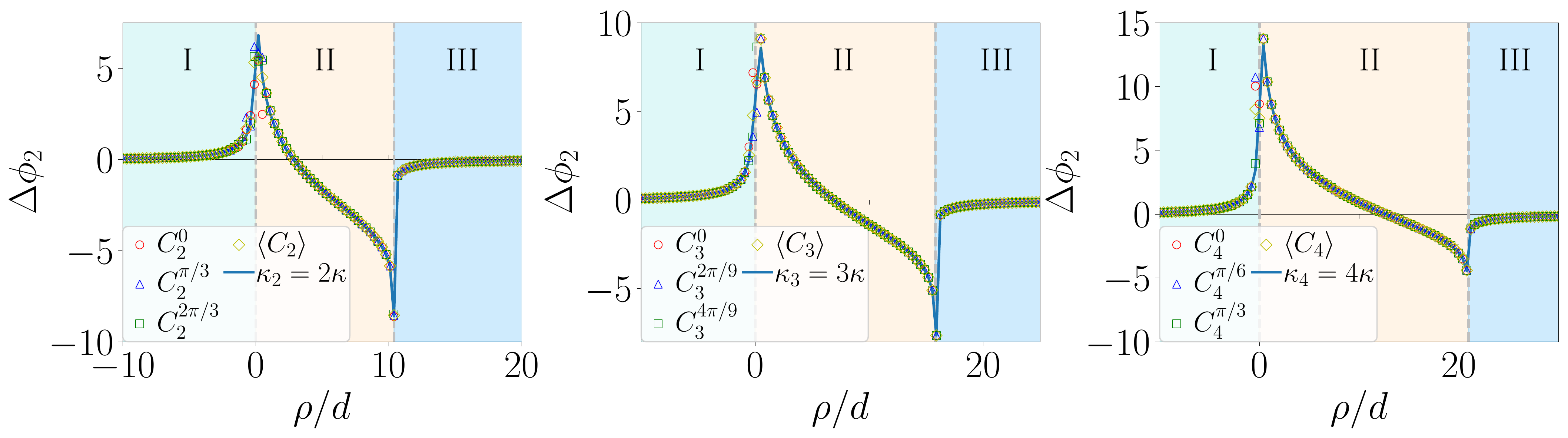}
	\caption{The dipole-cluster (markers indicating different phases) and dipole-$m\kappa$-vortex (solid blue curve) scattering angles compared for $m=2,3,4$. Regions I and III indicate direct scattering and region II pseudo-exchange scattering in the dipole-$m\kappa$-vortex simulations. Vertical gray dashed lines indicate numerically observed boundaries between the regions of interaction.
	\label{fig:18_cluster_angles}}
	\end{center}
\end{figure}

\noindent In figure~\ref{fig:18_cluster_angles} we plot the dipole-cluster and dipole-$m$-vortex scattering angles versus the normalized impact parameter $\rho/d$. Each subfigure displays the numerical data for the cluster $C_m$ with $m=2,3,4$ (left, middle, right) for three sets of cluster phases $\xi$, labeled by $C_m^\xi$, as well as the phase-averaged scattering angles $\langle C_m^\xi\rangle$. For $m=2$, we plot three phases uniformly distributed over $\xi\in[0,\pi)$, for $m=3$, we plot three phases uniformly distributed over $\xi\in[0,2\pi/3)$, and for $m=4$, we plot three phases uniformly distributed over $\xi\in[0,\pi/2)$. The scattering angles are also compared to those of the respective dipole-$m$-vortex simulations -- in the case for $m=3$ we exclude data if the dipole does not remain coherent. The background colors indicate the direct (regions I and III) and exchange (region II) scattering regions observed in the corresponding dipole-$m\kappa$-vortex simulations, whose scattering angles we also plot via the blue curves. We observe remarkable agreement between the scattering angles observed in the cluster and $m\kappa$-vortex simulations across all impact parameter values. Indeed, we observe only minor discrepancies close to the asymptotic region close to $\rho/d=0$ in a small band of impact parameter values of width  approximately $d$. It is in fact quite surprising that the agreement is so good even for $\rho/d$ close to $0$ where the vortex dipole collides head on towards the vortex cluster where we would expect that the extra degrees of freedom provided by the cluster would lead to more exotic interaction behavior. At the extremes of the impact parameter range presented we observe indistinguishable deviations between the cluster and $m\kappa$-vortex data confirming our hypothesis that for large impact parameters the three vortex description is an appropriate approximation.

What is most interesting in the pseudo-exchange interaction range is that the final scattering angle remains close to that of the dipole-$m\kappa$-vortex simulation for the same set of impact parameters even though the types of exchanges cannot occur in the latter case. Whilst a pure exchange scattering resulting in a new dipole is not technically possible in the dipole-$m\kappa$-vortex case (due to the mismatch of circulations), a partial exchange in the form of a pairing of the anti-vortex to the larger $m\kappa$-circulation vortex can occur in region II (not shown) resulting in the subsequent spiraling of the negative vortex back towards its original $\kappa$-circulation partner with a second exchange occurring resulting in the formation of the original dipole. This is reminiscent of the slingshot effect found in Price~\cite{price_chaotic_1993} and is very surprising that such complex dynamics is observed in both the dipole-$m\kappa$-vortex and dipole-cluster collisions. Ultimately, we observe a scattering angle picture that is qualitatively similar to the dipole-vortex interaction, with two asymptotes appearing at boundaries between different types of vortex interactions that correspond to infinite rotation of the anti-vortex. The small band of exchange interactions observed in figure~\ref{fig:17_cluster_interaction_regions} are all located close to the asymptote which makes it difficult for us to distinguish any unique behavior.  Moreover, we see little to no deviation with respect to change in the cluster phase close to $\rho/d=0$ where we would expect it to have a significant influence on the structure of the interaction.  

Subsequently, we can conclude that the $m\kappa$-circulation vortex approximates the $m$-cluster even better than first thought, with seemingly the only significant discrepancies occurring in the small region of parameter space where the propagating dipole directly collides with the cluster leading to complex dynamics that we cannot simulate by the dipole-$m$-vortex collision. Moreover, even in the latter cases the scatting angle characteristics remain robust except for rare ``reversed collapse" collisions leading to the splitting of the cluster and loss of the outgoing dipole.

\subsection{Final Dipole Size}

\begin{figure}[htp!]
	\begin{centering}
		\includegraphics[width = \textwidth]{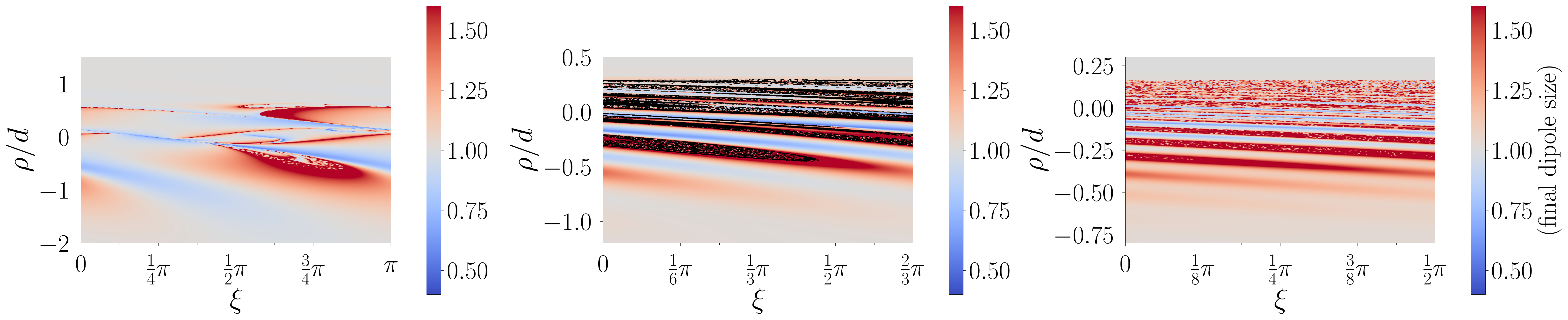}
	\caption{Final dipole separations normalized by the initial dipole separation $d$ of the dipole-cluster collisions with $m=2$ (left), $3$ (center), $4$ (right). Only for the $m=3$ cluster simulations do we observe disintegration of the vortex dipole and the vortex cluster indicated by regions colored in black. \label{fig:19_cluster_final_dipole_size}}
\end{centering}
\end{figure}

\noindent In figure~\ref{fig:19_cluster_final_dipole_size} we plot the final separation of the post-interaction vortex dipole in the dipole-cluster collisions that we consider. We measure the dipole separation post-interaction once the dipole has traversed a distance of at least $L+100d$ away from any vortex within the cluster. We observe similar banding that was seen in figure~\ref{fig:17_cluster_interaction_regions} across the same values of parameters. (For larger impact parameters -- not shown -- we simply observe that the final dipole distance relaxing back to $d$.) The fact that we only observe variability of the dipole size within a confined region close to $\rho/d=0$ is testament to the close approximation to the dipole-$m\kappa$-vortex interaction. This is because in the latter case the dipole cannot change size after propagating away for the same reason as the vortex-dipole interaction, due to the conservation of energy. This implies that changes in the final dipole size is due to the direct interaction of the dipole with the core of the cluster. Moreover, as the dipole size changes precisely in the same banding regions of the interaction type we see that the exchange interaction leads predominately to a widening of the dipole (regions colored red), while direct scattering leads to tightening of the dipole (regions colored blue). 

In the case of the three-cluster $C_3$ we note the possibility of the cluster breaking apart such as the trajectory found in figure~\ref{fig:16_cluster_example_trajectories} (bottom left). This means that there is no coherent dipole at later times that can be measured and hence in figure~\ref{fig:19_cluster_final_dipole_size} we color code these regions as black. This type of interaction was only observed for $C_3$ cluster interactions. We speculate that this may be the result of the odd number of vortices, as the disintegration of the $C_3$ cluster into two expanding $c_2$ clusters is approximate to the time-reversed vortex collapse solution of three vortices with circulations $(-2,-2,1)$ as found by Kudela~\cite{kudela_self-similar_2014}, it may be that a similar collapse solution does not exist for the other cluster collisions and is the reason why we did not observe similar behavior for $C_2$ and $C_4$. Across our simulations, we observe that in the $C_2$ case we have a maximum final dipole size of $6.18d$ and minimum of $0.35d$, in the $C_3$ case a maximum of $4.82d$ and a minimum of $0.15d$, and in the $C_4$ case a maximum of $4.16d$ and a minimum of $0.11d$. This indicates that the larger the cluster the more extreme the dipole vortex can shrink in size. 

Any change in the final dipole size must be compensated by a corresponding change in the cluster configuration in order to conserve the total energy of the system. In principle, an increase of the final dipole will result in the expanding of the vortex cluster and vice-versa.

With regards to figure~\ref{fig:18_cluster_angles} we have significant dipole scattering throughout a large portion of the parameter range. In figure~\ref{fig:19_cluster_final_dipole_size} however, the expansion or contraction of the dipole only occurs in a very small subsection of this region, and only in the region where the dipole-$m$-vortex approximation seems to breaks down during a head on collision. Moreover, it appears that the size of the region in which the dipole may change size after interaction decreases as the number of vortices in the cluster increases. For example, in the $C_2$ case the effective range for change in the dipole size is between $-1.5 < \rho/d < 0.6$ whereas in the 4-cluster case this is reduced to $-0.6 < \rho/d< 0.2$, reducing almost a third in size. This is perhaps due to the increased impulse on the dipole by the cluster reducing the ability of the internal structure of the cluster to have an effect. This narrowness is in contrast to the same phenomenon observed in the dipole-dipole interaction which occurred over a relatively large parameter range. In essence, we conclude that the dipole-dipole collision is much more effective at changing the dipole size than the assorted dipole-cluster collisions.

\begin{figure}[htp!]
	\begin{centering}
		\includegraphics[width = \textwidth]{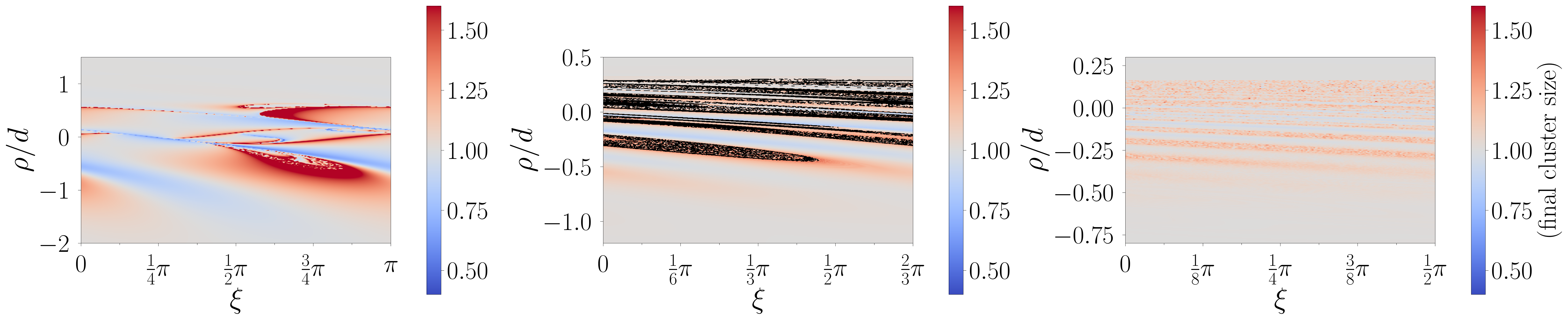}
		\caption{Final cluster separations normalized by the initial dipole separation $d$ of a dipole colliding with a 2,3, and 4 cluster respectively, shown as a heatmap against the general normalized coordinates $\rho/d$ and $L/d$. \label{fig:20_cluster_scattering_final_cluster_size}}
\end{centering}
\end{figure}

We define the cluster size $l_{C_m}$ by first computing the center of vorticity of the vortex cluster and then taking the average vortex separation distances of the vortices within the cluster to that center $ l_{C_m} = 1/m\sum_{i=1}^m \sqrt{(x_i - \bar{x})^2+(y_i - \bar{y})^2}$ where $\bar{x} = 1/m\sum_{i=1}^m x_i$, and $\bar{y} = 1/m\sum_{i=1}^m y_i$ with the summations taken only over the vortices within the cluster. In figure~\ref{fig:20_cluster_scattering_final_cluster_size} we plot a measure of the the final cluster size across the same parameter range as figure~\ref{fig:19_cluster_final_dipole_size}.

 We observe comparable results to those presented in figure~\ref{fig:19_cluster_final_dipole_size}. In the case of $C_2$ there is remarkable agreement between the final dipole and cluster sizes as would probably be expected as the cluster consists of the same number of vortices as a dipole. For $m>2$, the cluster has to contract or expand less to result in the equivalent energy offset that a contracting or expanding dipole would produce. This comes clearly from energy conservation as required by the Hamiltonian, as the product of the lengths between vortices in the cluster must increase proportionally to an increase in the size of the dipole, this means that as the final dipole size increases/decreases each individual length between cluster vortices has to increase/decrease less as the number of vortices in the cluster $m$ increases. This behavior is a characteristic of a dual cascade turbulent system. The vortex interaction can lead to a tightening of a vortex dipole which is an analogy of creating finer scale fluctuations (direct cascade of enstrophy) in a fluid flow. This is compensated by a more coherent vortex cluster at the largest scales (inverse cascade of energy). The fact that these two processes occur simultaneously is a principle outlined by Fj\o rtoft~\cite{fjortoft_changes_1953} for the development of a dual cascade between enstrophy and energy in 2D turbulence.

\begin{figure}[htp!]
	\begin{center}	
	\includegraphics[width = \textwidth]{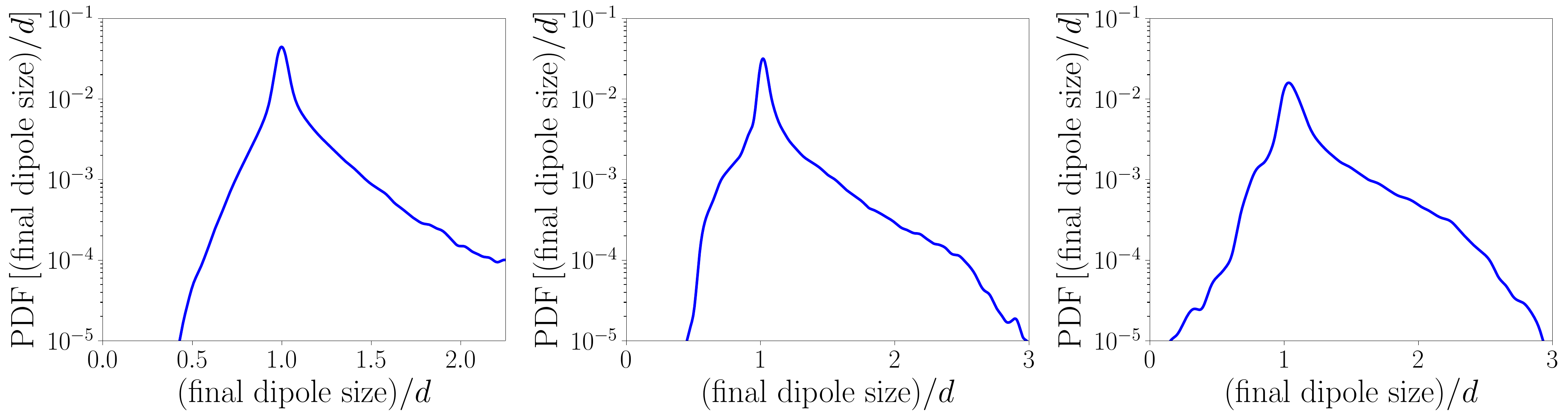}
	\caption{Probability density functions of the final dipole separations in the dipole-cluster interaction with $C_2$ (left), $C_3$ (middle), and $C_4$ (right). \label{fig:21_cluster_scattering_pdfs}}
\end{center}
\end{figure}

We additionally check the distribution of the final dipole sizes in the dipole-cluster collision by scanning over the parameter range. In figure~\ref{fig:21_cluster_scattering_pdfs} we present kernel density estimation to approximate the probability density distribution of the final dipole sizes, where in the case of $C_3$, we take only the numerical values where a coherent dipole remains post-interaction. We restrict the interval of impact parameters that we used as this removes a delta-function peak situated at size $d$ due to infinite states in which the dipole propagates past the vortex cluster at far distances. However, by averaging only across a range of impact parameters $-1.5 < \rho < 2$ in the $C_2$ case, $-1.2 < \rho < 0.5$ for $C_3$, $-0.8 < \rho < 0.3$ in the $C_4$ case we get the sense of the likelihood of increase or decreasing the dipole size. As first indicated in figure~\ref{fig:19_cluster_final_dipole_size} we see a propensity for the dipole to increase in size after interacting with larger clusters which is also compensated by the tendency for more smaller dipoles. We compute the expected dipole size post-interaction from our kernel density estimates  which gives $\langle\textrm{final dipole size}\rangle = 1.07d$ for $C_2$, $\langle\textrm{final dipole size}\rangle = 1.12d$ for $C_3$, and $\langle\textrm{final dipole size}\rangle = 1.19d$ for $C_4$. There is a slight increase in the mean as the cluster size grows, but in all cases the values are close to unity, and the likelihood that these values contain minor errors from the parameter space resolution of our numerical procedure is strong. Therefore, we cannot say for certain that these values indicate that the expected dipole distance to be significantly different from unity.

\section{Summary and Discussion} \label{sec:conclusion}

\noindent We have examined the interaction of a point vortex dipole with a variety of secondary vortex structures. These results provide important insight into the evolution of larger vortex systems: we hypothesize that in large vortex sets, dipole collisions will be the most fundamental interaction mechanism between vortex structures due to the ubiquity of dipoles across all temperature states and the fact that dipoles are the principle propagating structures. In section~\ref{sec:three_vortex} we presented the scattering angle results for the three-vortex dipole-vortex interaction and corrected an error from~\cite{aref_motion_1979} in his computation in region III.  We investigated the minimal and maximal dipole separation and showed that while a dipole cannot permanently change its size after interaction it can temporally grow or shrink for particular impact parameters. Significantly, we prove that the dipole-vortex interaction can lead to the dipole vortex shrinking down to a minimum size of $d/2$ attained when $\rho/d=-1$ at the critical point of the interaction. In fact temporary dipole shrinkage is found to  occur for parameters $\rho/d<-1$. This means for systems where the point vortex analogy makes sense, such as 2D BEC or optical turbulence, the three vortex interaction could provide a necessary process leading to vortex annihilation and a reduction in the overall vortex number. This vortex annihilation mechanism is of key importance when discussing other vortex tangle phenomena, such as vortex reconnections and sound emission \cite{ogawa_study_2002} with many works~\cite{nazarenko_freely_2007,salman_long-range_2016,baggaley_decay_2018} having highlighted the significant role that the vortex annihilation process has in the mixing of the turbulent states and in deciding the final decay states of the vortex systems. It is worth to note that the presence of sound in BECs has also been predicted to be a mechanism for the coming together of vortices and instigating vortex annihilation~\cite{nazarenko_freely_2007}, therefore it would be of interest to see which process is more efficient in large well-developed tangles. We leave this comparison as a future planned study.  

In section \ref{sec:int-4vortex} we expanded our study to dipole-dipole interactions. For the integrable case, we observed similar characteristics to the three vortex dipole-vortex interaction. This is due to the mathematical reduction to a three-body problem similar to the preceding three vortex interaction with the exception of the more complicated Hamiltonian expression accounting for the presence of the fourth vortex. We show qualitatively similar scattering behavior but with less pronounced temporary dipole size fluctuations compared to the interaction with a single vortex presumably due to the additional fourth vortex that damps the long-range interaction of the vortex structure. Interestingly, we highlighted several particularly interesting interactions, with one of these being related to the rotating Havelock ring configuration~\cite{havelock_stability_1931} observed when same-signed vortices collide on oppositely opposed trajectories, and a second being when the two dipoles interact head on mimicking  the interaction of a vortex ring with a solid boundary.

In section \ref{sec:non-int-4vortex} we considered non-integrable dipole-dipole configurations where we again saw the scattering (as in the integrable case) tends to lead to less dipole size fluctuations than the dipole-vortex collision. We also found that the non-integrable dipole-dipole interaction is the simplest collision in which persistent dipole growth or shrinkage is possible. We present a complex array of interactions that can lead to permanent changes in the final dipole size,  where the two final dipoles sizes $d_1, d_2$ must grow of shrink according to the relation $d^2=d_1d_2$ when infinitely separated due to energy conservation. This has an important physical consequence on the mixing of the vortex tangle and the momentum and energy exchange between two coherent vortex structures as it's the simplest example in which the vortex structures can permanently exchange these quantities. When concerned with finding the most effective route to vortex annihilation, one could conjecture a two-step scenario where the distance between the oppositely signed vortices is shrinking initially due to the non-integrable dipole-dipole collisions, which is followed by the final $d/2$ shrinking due to more frequent dipole-vortex scattering. 

Of course, for a large turbulent vortex flow, a broad variety of vortex structures other than dipoles and isolated vortices will exist. Therefore, it is prudent to consider what happens when dipoles interact with a  vortex cluster - we define cluster structures as consisting of more than two same-sign point vortices. One would expect that the most likely example of cluster interactions will be for a dipole to interact with these vortex clusters as dipoles are the primary structure propagating quickly across the vortex gas.  In section~\ref{sec:dipole-cluster} we consider isolated interactions between a point vortex dipole and a symmetric vortex cluster of sizes $m=2,3,4$. On their own, clusters of same-signed vortices will freely rotate around their center of vorticity, and in general (for rationally symmetric clusters) will not translate across the domain by themselves. Surprisingly, we showed for a significant majority of impact parameters, the dipole-cluster interactions can be effectively described by a three vortex interaction between a dipole and a third vortex of circulation $m\kappa$. From a far-field prospective this makes sense. At large distances the propagating dipole only experiences the mean effect of many same-sign vortices similar to a strong, but isolated, single vortex. In fact, even though we observe more complicated exchange and pseudo-exchange interactions between the dipole and cluster for relatively small impact parameters (this is due to the extra degrees of freedoms provided by the cluster), the final state, including the final dipole size and the observed dipole scattering angle, remain remarkably close to those of the three vortex analogue. Only in a very small band of impact parameters, that correspond to collisions where the dipole and cluster are very close at interaction, where we observe striated regions of direct and exchange scattering, do we observe any significant discrepancies. Furthermore, these are precisely in those parameter regimes where there is a significant scattering angle (rapid rotation of more than $2\pi$) and where the dipole propagates through the center of the cluster interacting with all of the cluster vortices. In the case of $C_3$ we observed cases of dipole/cluster disintegration leading to ever expanding mixed vortex structures. This is reminiscent of the time-reversed vortex collapse solutions observed by Kudela~\cite{kudela_self-similar_2014} in the case of three point vortices with circulations $(-2, -2, 1)$.  This particular collapse solution is interesting as in our simulation, the  cluster disintegrates, splitting into two 2-clusters and a lone negative vortex mimicking the vortex collapse solution (with opposite circulation signs). This poses questions of stability of the vortex collapse solution, because in our case, if we reverse time, we do not observe a collapse, but the formation of a vortex dipole and cluster, meaning that there is a possibility of ``near collapse" solutions that create coherent and stable vortex structures. The particular combination of vortex circulations of the Kudela collapse solution presumably accounts for the lack of cluster disintegration observed in the dipole interactions with a $C_2$ and $C_4$ clusters as there does not exist a way to form any vortex collapse solutions with the circulations in these interactions. Thus it is an interesting open problem whether there exists other vortex configurations that could replicate time-reversed vortex collapse solutions involving a higher number of point vortices as found numerically by Kudela~\cite{kudela_self-similar_2014}.  It would be interesting to know the role of the reversed process in which a tight vortex cluster and dipole quickly appear as a result of partial or frustrated vortex collapse.  

Overall, in the cluster interactions, the added degrees of freedom provided by the additional vortices means that there are little constraints imposed on the dipole in terms of changing size. However, we still observe that there is only a significant change when close to the direct and exchange interaction boundary close to $\rho/d =0$. Interestingly, most interactions, apart from a few exceptions just discussed, still lead to a coherent dipole and a vortex cluster. Any change in the dipole size, has to be accounted for by a like change in the overall cluster size due to the energy conservation when the two structures are infinitely apart. The corresponding change in the cluster size is then proportional to the number of vortices contained within it - as stipulated by the conservation of energy. (It required less shrinkage of a large cluster to offset any energy change in the dipole). This process has an interesting ``dual cascade" interpretation where the shrinking dipoles and the tightening clusters manifest an energy cascade to small scales and the enstrophy cascade to larger scales respectively. In table~\ref{tab:results_summary} we provide a summary of the results of our investigation into dipole sizes across the fundamental interactions that we consider. Note that dipole sizes during interaction are not given for the cluster $C_m$ cases due to the difficultly in defining a dipole distance when the dipole interacts with many vortices. We have investigated only the most basic vortex interactions with a dipole that are likely to appear in a fully developed turbulent vortex tangle, but we already see interesting dynamics and processes that could potentially have a significant impact of the topological properties and structure of a large tangle. We leave it as a future study to examine how common each type of process is but we imagine that the dipole-vortex, dipole-dipole and dipole-cluster may well be the most common for relatively random, uncorrelated, and dilute vortex systems. Furthermore, we have seen evidence that the vortex-cluster interactions appear similar to those of the three vortex dipole-$m\kappa$-vortex interaction that still remains an integrable system with gives hope for a potential theoretical treatment of vortex interactions using a kinetic theory. Finally, it would be interesting to investigate the situation of point vortex dynamics in the vicinity of a cluster of sufficiently large size that it could  be represented by a continuous large-scale mean flow. This could be studied in a kinetic framework similar to~\cite{nazarenko_kinetic_1992} where a gas of point vortices were evolving on a continuous shear flow background.

\begin{table}[]
	
    \begin{center}
        \caption{Summary of the dynamical quantities in the fundamental interactions in the point vortex model, with each column giving a different structure that a dipole may collide with and each row giving a different quantity of interest.\label{tab:results_summary}}
	\begingroup
        \renewcommand*{\arraystretch}{1.25}
	\resizebox{\textwidth}{!}{
    \begin{tabular}{c|c|c|c|c|c|c}
        \multirow{2}{*}{Configuration}&\multirow{2}{*}{Three Vortex}&\multicolumn{2}{c|}{Four vortex}&\multicolumn{3}{c}{Dipole-Cluster}\\  
         && Integrable & Non-Integrable & $C_2$ & $C_3$ & $C_4$  \\ \hline
        minimum dipole distance & $d/2$ &  $(d/\sqrt{2})\sqrt{\sqrt{5}-1}$    &$0.13d$&\rule{10pt}{0.7pt} & \rule{10pt}{0.7pt} & \rule{10pt}{0.7pt} \\ 
        maximum dipole distance & $2d$ & $\sqrt{2}d$ &$3.05d$&\rule{10pt}{0.7pt} & \rule{10pt}{0.7pt} & \rule{10pt}{0.7pt} \\
        maximum final dipole size & $d$ & $d$ & $2.88d$  & $6.18d$  & $4.82d$  & $4.16d$\\ 
        minimum final dipole size & $d$ & $d$ &  $0.016d$ &  $0.35d$ & $0.15d$  & $0.11d$\\ 
        $\langle\textrm{final dipole size}\rangle$ & $d$ & $d$ & $d$ & $1.07d$ & $1.12d$ & $1.19d$ \\ 
    \end{tabular}
    }
   \endgroup 
    \end{center} 
    
\end{table}

\section{Acknowledgments}
\noindent This work has received funding from the European Union’s Horizon 2020 research and innovation programme under the Marie Skłodowska-Curie grant agreement No 823937 for project HALT, and  the support of NVIDIA Corporation with the donation of the Titan Xp GPU used for this research.

\appendix

\section{Derivation of the $b_2$ evolution equation for the dipole-vortex collision \label{app:b2}}
 
We consider the three vortex interaction in terms of the dimensionless $b_i$ variables, derived from the $l_{ij}$ variables where $l_{ij}$ is the relative intervortex separation between vortex $i$ and vortex $j$, and we consider vortices with circulations $\kappa_1 = \kappa_2 = -\kappa_3 = \kappa$ (see dipole-vortex schematic in main article for more details)

\begin{equation}
b_1=\frac{l^2_{23}}{\kappa C}, \quad b_2=-\frac{l^2_{13}}{\kappa C}, \quad b_3=\frac{l^2_{12}}{\kappa C},
\end{equation}

\noindent and here the constant of motion $C$ is defined through the conserved quantity $R$:

\begin{equation}\label{eq:C}
R = \kappa^2l^2_{13}-\kappa^2l^2_{12}-\kappa^2l^2_{23} = -3\kappa^3C.
\end{equation}
This rescaling implies we have the constraint $b_1+b_2+b_3 = 3$ and one can interpret the variables $b_i$ for $i=1,2,3$ as tri-linear coordinates which must satisfy the geometric constraints
\begin{equation}
b_1^2 + b_2^2 +b _3^2 \leq 2(b_1b_3 -b_1b_2-b_2b_3 ),
\end{equation}
arising from triangle inequality. From the conservation of the Hamiltonian $H$, we can define a non-negative constant $\theta$ (as done by Aref~\cite{aref_motion_1979}) that will be useful in our analysis in the form

\begin{equation}
\frac{|b_2|}{b_1b_3} =  \kappa|C|\exp\left(-\frac{4\pi H}{\kappa^2}\right) = \theta.
\end{equation}
To examine the three vortex interaction we seek to find the equation of motion of the dimensionless variables $b_i$ for $i=1,2,3$. Of particular interest is the equation of motion for $b_2$ as it corresponds to the dynamics of the vortex separation $l_{13}$ of the two positive vortices which we can use as a proxy for the distance between the dipole and isolated vortex (the two positive vortices can never be part of the dipole together). This means we can encapsulate the dynamics of the collision using the solution for $b_2$. Using the evolution equation for the relative vortex motion

\begin{equation}
	\frac{dl_{ij}^{2}}{dt}=  \frac{2}{\pi}\sum_{k=1 , k\neq i ,  k \neq j}^{N}\kappa_{k}\epsilon_{ijk}A_{ijk}\left(\frac{1}{l_{jk}^{2}}-\frac{1}{l_{ki}^{2}}\right),
\end{equation}

\noindent with the definition of the dimensionless variables $b_1, b_2, b_3$, and the conservation laws of the point vortex interaction, one can show that
\begin{equation} \label{eq:A1}
\dot{b}_2=\left(-\frac{1}{C}\right)\frac{2}{\pi}\kappa_2\epsilon_{ijk} A_{123}\left(\frac{1}{l_{23}^2}-\frac{1}{l_{13}^2}\right)=\pm\frac{2}{C^2\pi}A_{123}\left(\frac{b_3-b_1}{b_1b_3}\right),
\end{equation}
where $A_{123}$ is the area of the triangle spanned by the three vortices and $\epsilon_{123}$ is the Levi-Civita symbol that indicates the orientation of the labeling of the triangle vertices. The area $A_{123}$ can also be expressed in terms of the vortex separations $l_{12}, l_{13}, l_{23}$ by Heron's formula $A_{123}=\sqrt{r(r-l_{12})(r-l_{23})(r-l_{13})}$ where $r=(1/2)(l_{12}+l_{13}+l_{23})$. This reduces to
\begin{equation*}
A_{123}=\frac{1}{4}\sqrt{4l_{12}^2l_{23}^2-\left(l_{12}^2-l_{13}^2+l_{23}^2\right)^2}.
\end{equation*}
By replacing with the dimensionless $b_1, b_2, b_3$ variables and using the constraints $b_1+b_2+b_3=3$ and the conserved quantity $\theta$ the area $A_{123}$ can be expressed in terms of variable $b_2$ only:  
\begin{equation*}
A_{123}=\frac{1}{4}\sqrt{4C^2b_1b_3-(Cb_3+Cb_2+Cb_1)^2}=\frac{C}{2\sqrt{\theta}}\sqrt{|b_2|-\frac{9}{4}\theta}.
\end{equation*}
In the same manner, the $(b_3-b_1)/b_1b_3$ term appearing in~(\ref{eq:A1}) can be expressed as 
\begin{eqnarray*}
	\frac{b_3-b_1}{b_1b_3}&=&\frac{\theta}{|b_2|}\sqrt{(b_3-b_1)^2}=\frac{\theta}{|b_2|}\sqrt{(b_1+b_2+b_3)^2-b_2^2-4b_1b_3-2b_1b_2-2b_2b_3},
			    \\&=&\frac{\theta}{|b_2|}\sqrt{(3-b_2)^2-4\frac{|b_2|}{\theta}}.
\end{eqnarray*}
Combining these results together gives us the dynamical equation for $b_2$ in terms of the variable $b_2$ alone

\begin{equation*}
 \dot{b}_2=\pm\frac{2}{C^2\pi}A_{123}\left(\frac{b_3-b_1}{b_1b_3}\right)=\pm\frac{\sqrt{\theta}}{C\pi b_2}\sqrt{(|b_2|-\frac{9}{4}\theta)\left[(3-b_2)^2-4\frac{|b_2|}{\theta}\right]}.
\end{equation*}

\begin{figure}[htp!]
	\begin{center}
		\includegraphics[width = \textwidth]{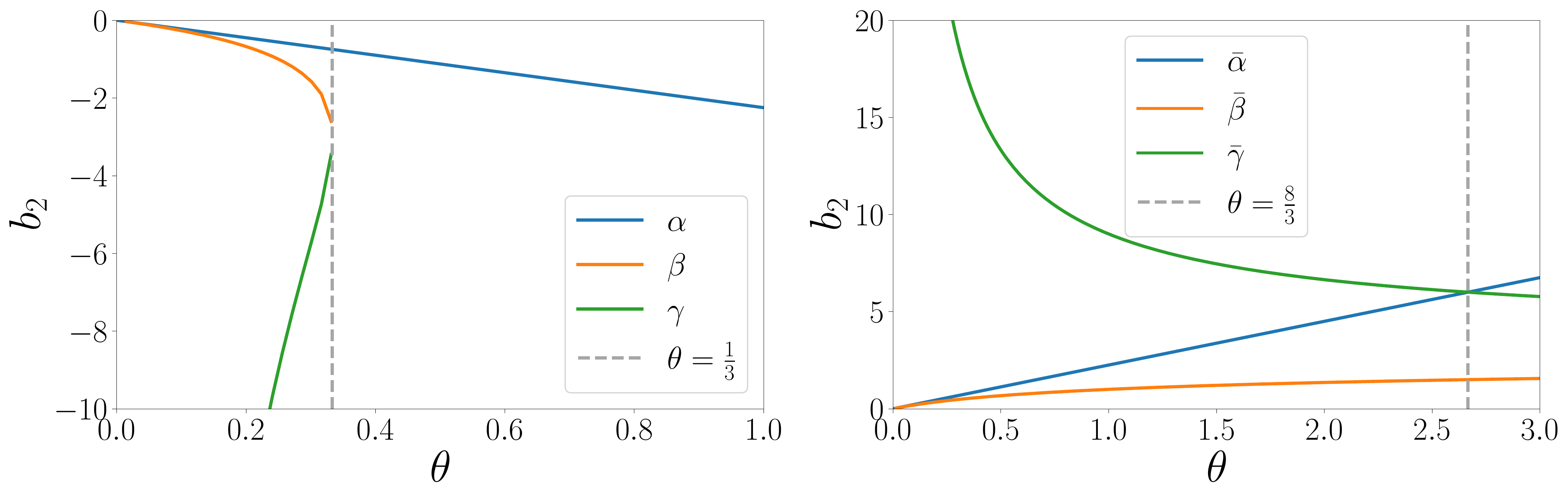}
	\caption{(Left) Plot of the roots $\alpha, \beta, \gamma$ for $C>0$ of~(\ref{eq:b2-2}). Roots $\beta$ and $\gamma$ become imaginary for $1/3 < \theta$. (Right) Roots $\bar{\alpha}, \bar{\beta}, \bar{\gamma}$ for $C<0$. All roots are positive, with $\bar{\gamma}$ is the largest root until $8/3 < \theta$ in which root $\bar{\alpha}$ becomes greatest.\label{fig:2_three_vortex_roots}}
	\end{center}
\end{figure}

By introducing the roots $\alpha, \beta, \gamma$ for $C>0$ and the roots $\bar{\alpha}, \bar{\beta},\bar{\gamma}$ for $C<0$ we attain the final form of $\dot{b}_2$

\begin{equation}\label{eq:b2-2}
	\dot{b}_2 =\begin{cases}\pm\frac{\sqrt{\theta}}{C\pi b_2}\sqrt{\left(\alpha-b_2\right)\left(\beta-b_2\right)\left(\gamma-b_2\right)} & \text{for $C>0$},\\
 \pm\frac{\sqrt{\theta}}{C\pi b_2}\sqrt{\left(b_2-\bar{\alpha}\right)\left(b_2-\bar{\beta}\right)\left(b_2-\bar{\gamma}\right)}& \text{for $C<0$},
\end{cases}
\end{equation}

\noindent these roots of $\dot{b}_2$ depend upon the sign of $C$ (or equivalently the sign of $b_2$ as it depends on $C$). These are, for $C>0$
\begin{equation}\label{eq:b2root_C<0}
\alpha(\theta)= -\frac{9}{4}\theta, \quad \beta(\theta) = -\frac{1}{\theta}\left(1-\sqrt{1-3\theta} \right)^2, \quad  \gamma(\theta) = -\frac{1}{\theta}\left(1+\sqrt{1-3\theta} \right)^2,
\end{equation}
and for $C<0$, the roots are
\begin{equation}\label{eq:b2root_C>0}
\bar{\alpha}(\theta)= \frac{9}{4}\theta, \quad  \bar{\beta}(\theta)= \frac{1}{\theta}\left(1-\sqrt{1+3\theta} \right)^2, \quad \bar{\gamma}(\theta) =  \frac{1}{\theta}\left(1+\sqrt{1+3\theta} \right)^2.
\end{equation}
The functional relationships of the above roots~(\ref{eq:b2root_C<0}) and~(\ref{eq:b2root_C>0}) are plotted verses $\theta$ in figure~\ref{fig:2_three_vortex_roots}. One should interpret the roots of $\dot{b_2}$ as follows: as $t\to-\infty$ the initial setup is of an isolated dipole infinitely far away from a third isolated point vortex. In the case of $C>0$, $b_2(-\infty)=-\infty$, while for $C<0$, $b_2(-\infty)=\infty$ where the value of $C$ depends on the initial configuration (and hence value of the impact parameter $\rho$). As the system evolves, the value of $b_2$ reduces continuously in magnitude (representing the dipole approaching the stationary vortex), this continues until it reaches the first root of $\dot{b}_2$ at which point it undergoes the corresponding scattering process. As can be determined from figure~\ref{fig:2_three_vortex_roots}, for $C>0$ and $\theta< 1/3$ the first root is $b_2=\gamma$ or for $C<0$ and $\theta < 8/3$, $b_2=\bar{\gamma}$  meaning that the condition $l_{12}=l_{23}$ is satisfied and the three vortex configuration undergoes an exchange scattering process. If the value of $\theta$ is $\theta>1/3$ for $C>0$ or $\theta>8/3$ for $C<0$ then the first critical point reached by $b_2$ is either $\alpha$ (for $C>0$) or $\bar{\alpha}$ (for $C<0$) respectively. At these values, the vortex configuration reaches a physical boundary equivalent to the limit of the triangle inequality, i.e. either $l_{23}=l_{12}+l_{13}$ for $C>0$ or $l_{13}=l_{12}+l_{23}$ for $C<0$ meaning that the three vortices are in a collinear configuration and a direct scattering process occurs. 

\section{Dipole-vortex scattering angles\label{app:aref}}

In the appendix of~\cite{aref_motion_1979} Aref computed the dipole-vortex scattering angles in the three vortex system. Here  we will re-derive the work of Aref due to is importance an relevance in the current manuscript, and in doing so, we will correct a mistake in the original derivation. We will utilize the evolution equation~(\ref{eq:b2}) (note that Aref used the labeling $b_3$) to determine the scattering angle of the vortex dipole, or specifically the negatively signed point vortex labeled as vortex $2$. The reason for tracking vortex $2$ is that this point vortex will always be part of the final vortex dipole pair being the only negatively signed point vortex. Therefore, we will use its direction of propagation as a proxy for the dipole scattering angle. Following Aref, we express the point vortex model in terms of polar coordinates $(r_i,\phi_i)$, emanating from the origin (or center of circulation), defined by $x_i = r_i\cos(\phi_i), \quad y_i = r_i\sin(\phi_i)$, where $r_i$ presents the distance of vortex $i$ from the center of circulation ${\bf x}_{\Gamma}$, and $\phi_i$ is the azimuthal angle. Then the Hamiltonian equation of motion of the point vortex system can be transformed into polar coordinates leading to
\begin{equation}\label{eq:ham_polar}
\kappa_ir_i\dot{r}_i = \frac{\partial H}{\partial \phi_i},\quad\kappa_ir_i\dot{\phi}_i=-\frac{\partial H}{\partial r_i}.
\end{equation}
Following the initial strategy outlined by Novikov~\cite{novikov_dynamics_1975}, we can transform our system from Cartesian coordinates to polar coordinates using the conserved quantities $H$ and $R$, and simple cosine laws for computing vortex separation around the center of circulation, with the result for the dynamics of $\dot{b}_2$. This leads to an evolution equation for $\dot{\phi}_2$ in terms of the variable $b_2$ giving
\begin{equation}\label{eq:phi_dot}
\dot{\phi}_2=\frac{\theta}{4\pi C}\frac{9-3b_2+\frac{4}{\theta}|b_2|}{(6-b
_2)|b_2|}.
\end{equation}
The scattering angle from the three vortex interaction can be subsequently defined by
\begin{equation*}
\Delta\phi_2=\lim_{t\to\infty}\phi_2(t)-\lim_{t\to-\infty}\phi_2(t)=\int^{\infty}_{-\infty}\ \dot{\phi}_2(t)\ dt.
\end{equation*}
Using~(\ref{eq:phi_dot}) and~(\ref{eq:b2}) for $\dot{\phi}_2$ and $\dot{b}_2$ enables us to define the scattering angle in terms of elliptic integrals with respect to $b_2$:

\begin{equation}\label{eq:delta_phi}
	\Delta\phi_2=\begin{cases}\mp\frac{\sqrt{\theta}}{4}\int^{L_2}_{L_1}\frac{9-(3+\frac{4}{\theta})b_2}{(6-b_2)\sqrt{(\alpha-b_2)(\beta-b_2)(\gamma-b_2)}}\ db_2 & \text{if $C>0$},\\
			\pm\frac{\sqrt{\theta}}{4}\int^{L_2}_{L_1}\frac{9-(3-\frac{4}{\theta})b_2}{(6-b_2)\sqrt{(b_2-\bar{\alpha})(b_2-\bar{\beta})(b_2-\bar{\gamma})}}\ db_2 & \text{if $C<0$.}
			\end{cases}
\end{equation}

\noindent where $L_1=b_2(t=-\infty)$, and $L_2=b_2(t=\infty)$ which will depend upon  the parameter $C$ and $\theta$.  The sign of the integrals will be determined by the sign of $\dot{b}_2$ during the evolution of $b_2$. We find that we have to consider the approach of the dipole pre-scattering separately from the evolution post-scattering. What we find is that the path taken by $b_2$ during the scattering process continuously evolves from the value of $L_1$ towards the first real root of the equation for $\dot{b}_2$. This root corresponds to the periapsis of $b_2$ in which we can formally define as the critical point. After which we must consider separately the evolution of $b_2$ towards $L_2$ (we shall see that this corresponds to a change of sign of $\dot{b}_2$. We can simplify the above result by considering the dipole-vortex setup in each of the scattering regions (I,II, III) separately. In each case this will specify the sign of $C$ and the integral limits $L_1$ and $L_2$ and enable us to perform the integral leading to an explicit formula for the scattering angle in terms of Legendre forms of elliptic integrals.

\subsection{Region I ($C >0$, $1/3 < \theta$)}

In region I, the impact parameter $\rho$ is large and positive and the scattering process will be of direct scattering type. For the values of the parameters $L_1=\lim_{t\to-\infty}b_2 = -\infty$. As the evolution progresses, the value of $b_2$ will increase until the first real root of $\dot{b}_2$ is reached. For the values of $C$ and $\theta$ in region I implies that roots $\beta$ and $\gamma$ are both complex meaning, so as $b_2$ must be real by definition, $b_2$ must increase from $-\infty$ until it reaches the physical boundary $\alpha$, whereupon it again rebounds back to $-\infty$. This implies that the scattering angle is given by the value of the integral:
\begin{eqnarray*}
	\Delta\phi_2&=&\frac{\sqrt{\theta}}{4}\int_{-\infty}^{\alpha}\frac{9-(3+\frac{4}{\theta})b_2}{(6-b_2)\sqrt{(\alpha-b_2)(\beta-b_2)(\gamma-b_2)}}\ db_2
	\\
		    &-&\frac{\sqrt{\theta}}{4}\int^{-\infty}_{\alpha}\frac{9-(3+\frac{4}{\theta})b_2}{(6-b_2)\sqrt{(\alpha-b_2)(\beta-b_2)(\gamma-b_2)}} \ db_2,
\\
		    &=&\frac{\sqrt{\theta}}{2}\int^{\alpha}_{-\infty}\frac{9-(3+\frac{4}{\theta})b_2}{(6-b_2)\sqrt{(\alpha-b_2)(\beta-b_2)(\gamma-b_2)}} \ db_2,
	\\ 
	db_2&=&\frac{\sqrt{\theta}}{2}\left[\left(3+\frac{4}{\theta}\right)\int^{\alpha}_{-\infty}\frac{1}{y}\ db_2+\left(9+\frac{24}{\theta}\right)\int^{\alpha}_{-\infty}\frac{1}{(b_2-6)y}\ db_2\right],
\end{eqnarray*}
where $y = \sqrt{(\alpha-b_2)(\beta-b_2)(\gamma-b_2)}$. Note that coefficients for the first expression are determined by whether each particular integral is regarding the stage of the interaction before or after scattering, e.g. in the first integral corresponds to $b_2$ from $-\infty$ to $\alpha$, and so we expect $\dot{b}_2>0$ hence this fixes the sign ($+$ in this case) arise in the $\dot{b}_2$ equation. The above integrals can be reduced to their Legendre Normal forms by simple substitutions found in Labahn and Mutrie~\cite{labahn_reduction_1997}. The Legendre forms for the first and third complete elliptic integrals are given by (note we are using the \textit{characteristic} $n$, given with an inverse sign than the usual third complete elliptic integral):
\begin{eqnarray*}
	K(k)&=&\int^{\pi/2}_{0}\frac{1}{\sqrt{1-k^2\sin^2\left(\mu\right)}}\ d\mu ,
\\
	\Pi(n,k)&=&\int^{\pi/2}_{0}\frac{1}{(n\sin^2\left(\mu\right)+1)\sqrt{1-k^2\sin^2\left(\mu\right)}}\ d\mu.
\end{eqnarray*}
To show this reduction, define a parameter $A=\sqrt{(\beta-\alpha)(\gamma-\alpha)}$ and split the first integral in the scattering angle expression into two with $b_2$ ranging from $-\infty$ to $\alpha-A$ and then $\alpha-A$ to $\alpha$,
\begin{equation*}
\int^{\alpha}_{-\infty}\frac{1}{y}\ db_3=\int^{\alpha-A}_{-\infty}\frac{1}{y}\ db_2+ \int^\alpha_{\alpha-A}\frac{1}{y}\ db_2,
\end{equation*}
and now consider the first integral on the right-hand side. Using the following substitutions
\begin{equation*}
	\sin^2\left(\mu\right) = \frac{8A(\alpha-b_2)}{(\alpha+A-b_2)^2}  ,\quad k^2=\frac{2A+2\alpha-\beta-\gamma}{4A},
\end{equation*}
this substitution gives a quadratic equation in $b_2$, the correct root depends upon whether the integral is from $-\infty\to\alpha-A$ or $\alpha-A\to\alpha$, using the positive root we can reduce the first integral to the following elliptic integral
\begin{equation*}
	\int^{\alpha-A}_{-\infty}\frac{1}{y}\ db_2=\int^{\pi/2}_{0}\frac{1}{\sqrt{A}\sqrt{1-k^2\sin^2\left(\mu\right)}}\ d\mu=\frac{1}{\sqrt{A}}K(k).
\end{equation*}

\noindent The same is done for the second part from $\alpha-A$ to $\alpha$, using the same expressions for $A$, $k$ and $\sin\left(\mu\right)$ and the negative root for $b_2$ leading to the second integral reducing to

\begin{equation*}
	\int^{\alpha}_{\alpha-A}\frac{db_2}{y}=\int^{0}_{\pi/2}\frac{-1}{\sqrt{A}\sqrt{1-k^2\sin^2\left(\mu\right)}}\ d\mu=\frac{1}{\sqrt{A}}K(k).
\end{equation*}

The final expression for the first integral becomes
\begin{equation*}
\int^{\alpha}_{-\infty}\frac{1}{y}\ db_2 =\int^{\alpha-A}_{-\infty}\frac{1}{y}\ db_2+\int^{\alpha}_{\alpha-A}\frac{1}{y}\ db_2 =\frac{2}{\sqrt{A}}K(k).
\end{equation*}
The second integral can be similarly reduced by splitting the integral into two parts: one with limits from $-\infty$ to $\alpha-A$ and a second with limits from $\alpha-A$ to $\alpha$ using the same substitutions as before with $n=-(\alpha+A-6)^2/[4A(\alpha-6) ]$, leading to

\begin{eqnarray*}
 &=&\int^{\alpha-A}_{-\infty}\frac{1}{(b_2-6)y}\ db_2+\int^{\alpha}_{\alpha-A}\frac{1}{(b_2-6)y}\ db_2\\	
						       &=&\frac{1}{\sqrt{A}}\int^{\pi/2}_{0}\frac{\sin^2\left(\mu\right)}{\left[(\alpha+A-6)\sin^2\left(\mu\right)+2A+2A\cos\left(\mu\right)\right]\sqrt{1-k^2\sin^2\left(\mu\right)}}\  d\mu
\\
						       &+&\frac{1}{\sqrt{A}}\int^{\pi/2}_{0}\frac{\sin^2\left(\mu\right)}{\left[(\alpha+A-6)\sin^2\left(\mu\right)+2A-2A\cos\left(\mu\right)\right]\sqrt{1-k^2\sin^2\left(\mu\right)}}\  d\mu,
\\
						       &=&\frac{2}{\sqrt{A}}\int^{\pi/2}_{0}\frac{(\alpha+A-6)\sin^2\left(\mu\right)-2A}{\left[\sin^2\left(\mu\right)(\alpha+A-6)^2-4A(\alpha-6)\right]\sqrt{1-k^2\sin^2\left(\mu\right)}}\ d\mu,
						     \\&=&\frac{2}{\sqrt{A}(\alpha+A-6)}\left[K(k)+\frac{1}{2}\left(\frac{A}{\alpha-6}-1\right)\Pi(n,k)\right].
\end{eqnarray*}

\noindent Subsequently, the full expression for the region I scattering angle can be simplified to
\begin{equation*}
\Delta\phi_3=\sqrt{\frac{\theta}{A}}\left[\left(3+\frac{4A}{\theta(\alpha+A-6)}\right)K(k)+\frac{\left(\alpha-A-6\right)}{\theta\left(\alpha+A-6\right)}\Pi(n,k)\right].
\end{equation*}

\subsection{Region IIa ($C>0$, $0 <\theta<1/3$)} 

Region IIa corresponds the case where $C>0$ and $0 <\theta<1/3$ leading to exchange scattering and as such the evolution of the variable $b_2$ will evolve from the $t\to-\infty$ limit $-\infty$ until it reaches its periapsis at the root $b_2=\gamma$ where the vortex interaction reaches the configuration with $b_1=b_3$. At this point, an exchange interaction occurs and $b_2$ decreases towards $-\infty$ again (see phase point diagrams in~\cite{aref_motion_1979} for clarification of this). Therefore, the scattering angle for vortex $2$ becomes
\begin{eqnarray*}
	\Delta\phi_2&=&\frac{\sqrt{\theta}}{4}\int_{-\infty}^{\gamma}\frac{9-(3+\frac{4}{\theta})b_2}{(6-b_2)\sqrt{(\alpha-b_2)(\beta-b_2)(\gamma-b_2)}}\ db_2
	\\
		    &-&\frac{\sqrt{\theta}}{4}\int^{-\infty}_{\gamma}\frac{9-(3+\frac{4}{\theta})b_2}{(6-b_2)\sqrt{(\alpha-b_2)(\beta-b_2)(\gamma-b_2)}}\ db_2,
\\&=&\frac{\sqrt{\theta}}{2}\int^{\gamma}_{-\infty}\frac{9-(3+\frac{4}{\theta})b_2}{(6-b_2)\sqrt{(\alpha-b_2)(\beta-b_2)(\gamma-b_2)}}\ db_2
\\
  &=&\frac{\sqrt{\theta}}{2}\left[\left(3+\frac{4}{\theta}\right)\int^{\gamma}_{-\infty}\frac{1}{y}\ db_2 +\left(9+\frac{24}{\theta}\right)\int^{\gamma}_{-\infty}\frac{1}{(b_2-6)y}\ db_2\right].
\end{eqnarray*}
again where we have defined the variable $y = \sqrt{(\alpha-b_2)(\beta-b_2)(\gamma-b_2)}$. To reduce the above formula into normal form, we begin by applying the substitution
 \begin{equation*}
	\sin^2\left(\mu\right)= \frac{\alpha-\gamma}{\alpha - b_2}, \qquad k^2=\frac{\alpha-\beta}{\alpha-\gamma},
\end{equation*}
to both integrals. Then the first integral becomes
\begin{equation*}
	\int^{\gamma}_{-\infty}\frac{1}{y}\ db_2=\frac{2}{\sqrt{\alpha-\gamma}}K(k),
\end{equation*}
while the second integral becomes
\begin{eqnarray*}
	&=&\frac{2}{\sqrt{\alpha-\gamma}}\int^{\pi/2}_{0}\frac{\sin^2\left(\mu\right) }{[(\alpha-6)\sin^2\left(\mu\right)-(\alpha-\gamma)]\sqrt{1-k^2\sin^2\left(\mu\right)}}\ d\mu,\\
						       &=& \frac{2}{(\alpha-6)\sqrt{\alpha-\gamma}}\left[K(k)-  \int_0^{\pi/2} \frac{1}{\left[ n\sin^2\left(\mu\right)+1\right]\sqrt{1-k^2\sin^2\left(\mu\right)}} \ d\mu\right] 
						     \\&=&\frac{2}{(\alpha-6)\sqrt{\alpha-\gamma}}\left[K(k)-\Pi(n,k)\right],
\end{eqnarray*}
where the parameter $n=(\alpha-6)/(\gamma-\alpha)$. Finally, returning to the full expression for the scattering angle for region IIa, we have  
\begin{eqnarray*}
	\Delta\phi_2&=&\sqrt{\frac{\theta}{\alpha-\gamma}}\left[\left(\frac{3\alpha+\frac{4\alpha}{\theta}-9}{\alpha-6}\right)K(k)-\left(\frac{9+\frac{24}{\theta}}{\alpha-6}\right)\Pi(n,k)\right] 
	\\
		    &=&\sqrt{\frac{\theta}{\alpha-\gamma}}\left[3K(k)+\frac{4}{\theta}\Pi(n,k)\right] .
\end{eqnarray*}

\subsection{Region IIb ($C<0$,  $ 0 < \theta<8/3$)}

For region IIb,  $C<0$ with $0 < \theta<8/3$ and therefore at the initial condition when the dipole is far from the isolated point vortex, the variable tends towards $b_2\to\infty$ as $t\to \infty$. As the vortex system evolves, then $b_2$ reduces until it reaches it periapsis at the largest root of the $\dot{b}_2$ equation for this parameter region, i.e. $b_2=\bar{\gamma}$. Once at the periapsis point, the positive vortices exchange in the dipole and the dipole propagates away with the variable $b_2$ increasing back towards $\infty$. Therefore, initially $b_2$ is decreasing (hence $\dot{b}_2<0$) pre-scattering, while post-scattering we have $\dot{b}_2>0$. This defines the signs to take in the scattering angle formula:
\begin{eqnarray*}
	\Delta\phi_2&=&-\int^{\bar{\gamma}}_{\infty}\frac{9-(3-\frac{4}{\theta})b_2}{(6-b_2)\sqrt{\strut(b_2-\bar{\alpha})(b_2-\bar{\beta})(b_2-\bar{\gamma})}}\ db_2
	\\
		    &+&\frac{\sqrt{\theta}}{4}\int^{\infty}_{\bar{\gamma}}\frac{9-(3-\frac{4}{\theta})b_2}{(6-b_2)\sqrt{\strut(b_2-\bar{\alpha})(b_2-\bar{\beta})(b_2-\bar{\gamma})}}\ db_2\\
		    &=&\frac{\sqrt{\theta}}{2}\int_{\bar{\gamma}}^{\infty}\frac{9-(3-\frac{4}{\theta})b_2}{(6-b_2)\sqrt{\strut(b_2-\bar{\alpha})(b_2-\bar{\beta})(b_2-\bar{\gamma})}}\ db_2,
		    \\
		    &=&\frac{\sqrt{\theta}}{2}\left[\left(3-\frac{4}{\theta}\right)\int^{\infty}_{\bar{\gamma}}\frac{db_2}{y}+\left(9-\frac{24}{\theta}\right)\int^{\infty}_{\bar{\gamma}}\frac{db_2}{(b_2-6)y}\right],
\end{eqnarray*}

\noindent where again we have defined a variable $y=\sqrt{\left(b_2-\bar{\alpha}\right)\left(b_2-\bar{\beta}\right)\left(b_2-\bar{\gamma}\right)}$. We reduce both integrals in the same way as before, using the substitutions
\begin{equation*}
	\sin^2\left(\mu\right) = \frac{b_2-\bar{\gamma}}{b_2-\bar{\alpha}},\quad k^2=\frac{\bar{\alpha}-\bar{\beta}}{\bar{\gamma}-\bar{\beta}},
\end{equation*}
The first integral subsequently becomes
\begin{equation*}
\int^{\infty}_{\bar{\gamma}}\frac{1}{y}\ db_2=\frac{2}{\sqrt{\strut\bar{\gamma}-\bar{\beta}}}K(k),
\end{equation*}
and the second integral becomes 
\begin{eqnarray*}
	\int^{\infty}_{\bar{\gamma}}\frac{1}{(b_2-6)y}\ db_2&=&\frac{2}{\sqrt{\strut\bar{\gamma}-\bar{\beta}}}\int^{\pi/2}_{0}\frac{\sin^2\left(\mu\right)-1}{\left[ (\bar{\alpha}-6)\sin^2\left(\mu\right)+6-\bar{\gamma}\right ]\sqrt{1-k^2\sin^2\left(\mu\right)}}\ d\mu\\
							    & = &\frac{2}{(6-\bar{\gamma})\sqrt{\strut\bar{\gamma}-\bar{\beta}}}\int^{\pi/2}_{0}\frac{\sin^2\left(\mu\right)-1}{[n\sin^2\left(\mu\right)+1]\sqrt{1-k^2\sin^2\left(\mu\right)}}\ d\mu,
\\
							    &=&\frac{2}{\left(\bar{\alpha}-6\right)\sqrt{\strut\bar{\gamma}-\bar{\beta}}}\left[K(k)-\frac{\bar{\alpha}-\bar{\gamma}}{6-\bar{\gamma}}\Pi(n,k)\right],
\end{eqnarray*}

\noindent where we have introduced a new parameter $n=(\bar{\alpha}-6)/( 6-\bar{\gamma} )$. Then the final  full expression for the  scattering angle in region IIb is
\begin{eqnarray*}
\Delta\phi_2=\sqrt{\frac{\theta}{\bar{\gamma}-\bar{\beta}}}\left[\left(\frac{3\bar{\alpha}-\frac{4\bar{\alpha}}{\theta}-9}{\bar{\alpha}-6}\right)K(k)-\left(\frac{\left(9-\frac{24}{\theta}\right)(\bar{\alpha}-\bar{\gamma})}{(\bar{\alpha}-6)(6-\bar{\gamma})}\right)\Pi(n,k)\right]
\\
=\sqrt{\frac{\theta}{\bar{\gamma}-\bar{\beta}}}\left[3K(k)-\frac{4}{\theta}\frac{(\bar{\alpha}-\bar{\gamma})}{(6-\bar{\gamma})}\Pi(n,k)\right].
\end{eqnarray*}

\subsection{Region III ($C < 0$, $8/3< \theta$)}

For region III, the final case corresponding to direct scattering with $C < 0$ and $8/3< \theta$, we have that in the limit of $t\to-\infty$, $b_2\to \infty$, with $b_2$ initially decreasing until it reaches the periapsis point of $b_2=\bar{\alpha}$, where the sign of $\dot{b}_2$ changes post-scattering and increases back up towards $\infty$. Therefore, the scattering angle integral~(\ref{eq:delta_phi}) becomes (here the sign of each integral is determined by the sign of $\dot{b}_2$ pre- and post-scattering)
\begin{eqnarray*}
	\Delta\phi_2&=&\frac{\sqrt{\theta}}{4}-\int^{\bar{\alpha}}_{\infty}\frac{9-(3-\frac{4}{\theta})b_2}{(6-b_2)\sqrt{\strut(b_2-\bar{\alpha})(b_2-\bar{\beta})(b_2-\bar{\gamma})}}\ db_2
	\\
		    &+&\frac{\sqrt{\theta}}{4}\int^{\infty}_{\bar{\alpha}}\frac{9-(3-\frac{4}{\theta})b_2}{(6-b_2)\sqrt{\strut(b_2-\bar{\alpha})(b_2-\bar{\beta})(b_2-\bar{\gamma})}}\ db_2, 
		  \\&=&\frac{\sqrt{\theta}}{2}\int_{\bar{\alpha}}^{\infty}\frac{9-(3-\frac{4}{\theta})b_2}{(6-b_2)\sqrt{\strut(b_2-\bar{\alpha})(b_2-\bar{\beta})(b_2-\bar{\gamma})}}\ db_2
		  \\
		    &=&\frac{\sqrt{\theta}}{2}\left[\left(3-\frac{4}{\theta}\right)\int^{\infty}_{\bar{\alpha}}\frac{db_2}{y}+\left(9-\frac{24}{\theta}\right)\int^{\infty}_{\bar{\alpha}}\frac{db_2}{(b_2-6)y}\right],
\end{eqnarray*}
with $y=\sqrt{\strut(b_2-\bar{\alpha})(b_2-\bar{\beta})(b_2-\bar{\gamma})}$. The substitution in this case is given as:
\begin{equation*}
	\sin^2\left(\mu\right) = \frac{b_3-\bar{\alpha}}{b_3-\bar{\gamma}},\quad k^2=\frac{\bar{\gamma}-\bar{\beta}}{\bar{\alpha}-\bar{\beta}},
\end{equation*}
which leads to the integrals simplifying to
\begin{equation*}
	\int^{\infty}_{\bar{\alpha}}\frac{1}{y}\ db_2= \int^{\pi/2}_{0}\frac{2}{\sqrt{\strut\bar{\alpha}-\bar{\beta}}\sqrt{1-k^2\sin^2\left(\mu\right)}}\ d\mu = \frac{2}{\sqrt{\strut\bar{\alpha}-\bar{\beta}}}K(k).
\end{equation*}
and
\begin{eqnarray*}
	\int^{\infty}_{\bar{\alpha}}\frac{1}{(b_3-6)y}\ db_3&=&\frac{2}{\sqrt{\strut\bar{\alpha}-\bar{\beta}}}\int^{\pi/2}_{0}\frac{\sin^2\left(\mu\right)-1}{\left[\left(\bar{\gamma}-6\right)\sin^2\left(\mu\right)+6-\bar{\alpha}\right]\sqrt{1-k^2\sin^2\left(\mu\right)}}\ d\mu,
							  \\&=&\frac{2}{(\bar{\gamma}-6)\sqrt{\strut\bar{\alpha}-\bar{\beta}}}\left[K(k)-\frac{\bar{\gamma}-\bar{\alpha}}{6-\bar{\alpha}}\Pi(n,k)\right].
\end{eqnarray*}
where we have defined the parameter $n=(\bar{\gamma}-6)/(6-\bar{\alpha})$. Subsequently, the final scattering angle expression in Legendre normal form is given as 
\begin{equation*}
	\Delta\phi_2=\frac{\sqrt{\theta}}{(\bar{\gamma}-6)\sqrt{\strut\bar{\alpha}-\bar{\beta}}}\left[ \left(\left[3-\frac{4}{\theta}\right]\bar{\gamma}-9\right)K(k)+\frac{4\left(\bar{\gamma}-\bar{\alpha}\right)}{\theta}\Pi(n,k)\right].
\end{equation*}

\subsection{Summary of the dipole-vortex scatting angle normal form reduction}

In summary, the dipole-vortex scattering angle calculation is given by the solution to~(\ref{eq:delta_phi}), where the path along which the integral is taken is defined by the values of variable $L_1= b_2(t\to-\infty)$ and $L_2= b_2(t\to \infty)$ via the periapsis of $b_2$. This means in~(\ref{eq:delta_phi}), we must consider the approach and departure of the dipole separately. For the approach, we must use $\dot{b}_2$ with the appropriate sign corresponding to the sign $\dot{b}_2$ for our region, while after scattering the angle will be determined by~(\ref{eq:delta_phi}) using the other sign of $\dot{b}_2$ as in all cases the value of $b_2(t\to-\infty) = b_2(t\to\infty)$. For each case we have shown that the scattering angle formulae correspond to the following integrals:

\begin{equation}\label{eq:scattering_angles}
	\Delta\phi_2=\begin{cases}
\displaystyle\frac{\sqrt{\theta}}{2}\int^{\alpha}_{-\infty}\frac{9-\left(3+\frac{4}{\theta}\right)b_2}{(6-b_2)\sqrt{(\alpha-b_2)(\beta-b_2)(\gamma-b_2)}} \ db_2 & \text{in region I},\\
 \displaystyle\frac{\sqrt{\theta}}{2}\int^{\gamma}_{-\infty}\frac{9-\left(3+\frac{4}{\theta}\right)b_2}{(6-b_2)\sqrt{(\alpha-b_2)(\beta-b_2)(\gamma-b_2)}}\ db_2& \text{in region IIa},\\
\displaystyle  \frac{\sqrt{\theta}}{2}\int_{\bar{\gamma}}^{\infty}\frac{9-\left(3-\frac{4}{\theta}\right)b_2}{(6-b_2)\sqrt{\strut(b_2-\bar{\alpha})(b_2-\bar{\beta})(b_2-\bar{\gamma})}} \ db_2& \text{in region IIb},\\
\displaystyle   \frac{\sqrt{\theta}}{2}\int_{\bar{\alpha}}^{\infty}\frac{9-\left(3-\frac{4}{\theta}\right)b_2}{(6-b_2)\sqrt{\strut(b_2-\bar{\alpha})(b_2-\bar{\beta})(b_2-\bar{\gamma})}}\ db_2 & \text{in region III.}
\end{cases}
\end{equation}

\noindent These can then be further simplified in the form of linear combinations of the Legendre complete forms of elliptic integrals.
\begin{equation}\label{eq:scatexpr}
 \Delta\phi_2=\sqrt{a_1}\left[a_2K(k)+a_3\Pi(n,k)\right],
 \end{equation}
where each constant coefficient $a_1, a_2, a_3$ and parameters $k$ and $n$ are given in table~\ref{tab:3-vortex-cofficients}. 

\begin{table}[htp!]
\begin{center}
	\caption{Coefficients $a_1, a_2, a_3$ and parameters $k$ and $n$ for the dipole-vortex scattering angle are presented by region of interaction. The full scattering angle of the integrable case is determined from the Legendre normal form~(\ref{eq:scatexpr}). Note that in Region IIa we have defined an auxiliary  parameter $A^2=(\gamma-\alpha)(\beta-\alpha)$.\label{tab:3-vortex-cofficients}}
\setlength{\tabcolsep}{5pt}
{\renewcommand{\arraystretch}{2.5}
\begin{tabular}{c|ccccc}
Region &$a_1$&$a_2$&$a_3$&$k^2$&$n$\\
\hline
I&$\frac{\theta}{\alpha-\gamma}$&$3$&$\frac{4}{\theta}$&$\frac{\alpha-\beta}{\alpha-\gamma}$&$\frac{6-\alpha}{\alpha-\gamma}$\\

IIa&$\frac{\theta}{A}$&$3+\frac{4}{\theta}+\frac{9+(24/\theta)}{\alpha+A-6}$&$\frac{\left[A/(\alpha-6)-1\right]\left[9+(24/\theta)\right]}{2(\alpha+A-6)}$&$\frac{A-(\beta+\gamma)/2+\alpha}{2A}$&$-\frac{(\alpha+A-6)^2}{4A(\alpha-6)}$\\

IIb&$\frac{\theta}{\bar{\gamma}-\bar{\beta}}$&$3$&$-\frac{\left[9-(24/\theta)\right]\left[\bar{\alpha}-\bar{\gamma}\right]}{(\bar{\alpha}-6)(6-\bar{\gamma})}$&$\frac{\bar{\alpha}-\bar{\beta}}{\bar{\gamma}-\bar{\beta}}$&$\frac{\bar{\alpha}-6}{6-\bar{\gamma}}$\\

III&$\frac{\theta}{\bar{\alpha}-\bar{\beta}}$&$\frac{3\bar{\gamma}-(4\bar{\gamma}/\theta)-9}{\bar{\gamma}-6}$&$-\frac{\left[9-(24/\theta)\right]\left[\bar{\gamma}-\bar{\alpha}\right]}{(\bar{\gamma}-6)(6-\bar{\alpha})}$&$\frac{\bar{\gamma}-\bar{\beta}}{\bar{\alpha}-\bar{\beta}}$&$\frac{\bar{\gamma}-6}{6-\bar{\alpha}}$
\end{tabular}
}
\end{center}
\end{table}  


 \section{Derivation of the $b_2$ evolution equation for the integrable dipole-dipole collision \label{app:four_vortex_b2}}

 To express the equation of motion for the integrable four-vortex case we again use the evolution equation of relative vortex motion , which is a general equation for any number of point vortices, we have 

\begin{eqnarray}\label{eq:4vortex_eq_motion}
	\frac{dl^2_{13}}{dt}&=&\frac{2}{\pi}\kappa_{2}\epsilon_{132}A_{132}\left(\frac{1}{l_{23}^2}-\frac{1}{l_{12}^2}\right)+\frac{2}{\pi}\kappa_{4}\epsilon_{134}A_{134}\left(\frac{1}{l_{34}^2}-\frac{1}{l_{14}^{2}}\right)\nonumber
	\\
			    &=&\pm\frac{2\kappa}{\pi}\left[A_{132}\left(\frac{1}{l_{23}^2}-\frac{1}{l_{12}^2}\right)-A_{134}\left(\frac{1}{l_{34}^2}-\frac{1}{l_{14}^2}\right)\right]\nonumber \\
			    &=&\pm\frac{4\kappa}{\pi}A_{123}\left(\frac{1}{l_{23}^2}-\frac{1}{l_{12}^2}\right),
\end{eqnarray}
where we have used $\kappa_2 = \kappa_4=-\kappa$, the fact that $\epsilon_{132}$ and $\epsilon_{134}$ will always be of opposite sign and the geometry of the parallelogram that implies that $A_{132}$ and $A_{134}$ are congruent triangles. As was similar to the dipole-vortex collision, critical points of~(\ref{eq:4vortex_eq_motion}) for variable $l_{13}$ occur when the vortex separations $l_{12}=l_{23}$ or if the three vortices $1, 2, 3$ leading to the area $A_{123}=0$. Following the same strategy that we outlined for the three vortex interaction, it is possible to form an analogous conserved quantity $\theta$ defined by

\begin{equation}
	\frac{\sqrt{b_2(b_2-6)}}{b_1b_3} = \kappa |C|\exp\left(-\frac{2\pi H}{\kappa^2}\right) = \theta,
\end{equation}

\noindent we also define $C$ from the three vortex $R$ as is given in~(\ref{eq:C}). Using the conserved quantities $C$ and $\theta$  and non-dimensionalizing~(\ref{eq:4vortex_eq_motion}) in terms of the dimensionless variable $b_2$ gives

\begin{equation}\label{eq:4vortex_eq_dimensionless-2}
	\frac{db_2}{dt}=\begin{cases}\frac{2\sqrt{\theta}}{C\pi\sqrt{b_2(b_2-6)}}\sqrt{(b_2-\alpha)(b_2-\beta)(b_2-\gamma)} & \text{if $C>0$},\\
\frac{2\sqrt{\theta}}{C\pi\sqrt{b_2(b_2-6)}}\sqrt{(\bar{\alpha}-b_2)(\bar{\beta}-b_2)(\bar{\gamma}-b_2)} & \text{if $C<0$.}
\end{cases}
\end{equation}

\noindent We have defined the roots of~(\ref{eq:4vortex_eq_dimensionless-2}) as the following: for the case when $C>0$
\begin{eqnarray}
	\alpha&=&3\left(1-\sqrt{1+\frac{9}{16}\theta^2}\right),\label{eq:4vortex_roots}
	\\
	\beta&=&3\left(1-\frac{2\sqrt{2}}{3\theta}\sqrt{1-\sqrt{1-\frac{9}{4}\theta^2}}\right),\nonumber 
	\\
	\gamma&=&3\left(1-\frac{2\sqrt{2}}{3\theta}\sqrt{1+\sqrt{1-\frac{9}{4}\theta^2}}\right),\nonumber
\end{eqnarray}
and when $C<0$ the roots are given by
\begin{eqnarray}\label{eq:4vortex_cneg_roots}
	\bar{\alpha}&=&3\left(1+\sqrt{1+\frac{9}{16}\theta^2}\right),
	\\
	\bar{\beta}&=&3\left(1+\frac{2\sqrt{2}}{3\theta}\sqrt{1-\sqrt{1-\frac{9}{4}\theta^2}}\right), \nonumber
	\\
	\bar{\gamma}&=&3\left(1+\frac{2\sqrt{2}}{3\theta}\sqrt{1+\sqrt{1-\frac{9}{4}\theta^2}}\right).\nonumber
\end{eqnarray}

\begin{figure}[htp!]
	\begin{center}
		\includegraphics[width = \textwidth]{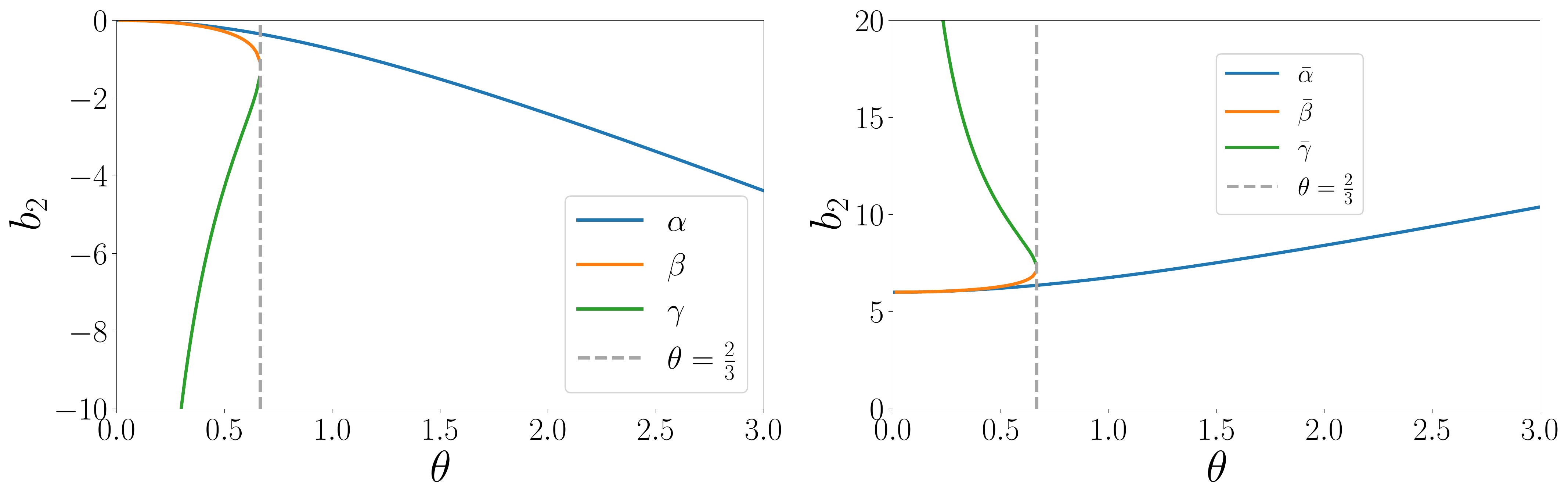}
	\caption{(Left) Plot of the roots $\alpha, \beta, \gamma$ for $C>0$ of~(\ref{eq:4vortex_eq_dimensionless-2}). Roots $\beta$ and $\gamma$ become imaginary for $2/3 < \theta$. (Right) Roots $\bar{\alpha}, \bar{\beta}, \bar{\gamma}$ for $C<0$ which are the mirror images of $\alpha, \beta, \gamma$ in the $b_2=3$ plane. Roots $\bar{\beta}$ and $\bar{\gamma}$ become imaginary for $2/3 < \theta$. \label{fig:7_four_vortex_roots}}
	\end{center}
\end{figure}

The boundaries between different scattering regimes (direct and exchange) in the dipole-dipole interaction can be determined as in the case of the vortex-dipole collision and are dictated by the roots of~(\ref{eq:4vortex_eq_dimensionless-2}) explicitly given by~(\ref{eq:4vortex_roots}) and~(\ref{eq:4vortex_cneg_roots}) and plotted in figure~\ref{fig:7_four_vortex_roots}. 

The symmetry of the two sets of roots around $b_2=3$ for either sign of $C$ is due to the parallelogram symmetry of the problem and is confirmed in the phase diagrams of Aref~\cite{aref_four-vortex_1999}.
For $\theta < 2/3$ we have three real roots in each case (dependent on the sign of $C$). Given the initial conditions where we assume $L\to \infty$ we have that  the first root attained by the variable $b_2$ during its evolution are $\gamma$ and $\bar{\gamma}$ for $C>0$ and $C<0$ respectively. These roots correspond to exchange scattering where the positive vortices are exchanges within the two vortex dipoles. For $2/3 < \theta$, the roots $\beta, \bar{\beta}$ and $\gamma, \bar{\gamma}$ become complex, see figure~\ref{fig:7_four_vortex_roots} for a graphical interpretation, and thus, the dipole interaction type is governed solely by the roots $\alpha$ and $\bar{\alpha}$ for $C>0$ and $C<0$ respectively that correspond to regions of direct scattering in the three vortex reduction where the three vortices become collinear.

\bibliographystyle{unsrt}
\bibliography{jason.bib}

\end{document}